\documentclass[]{article}
\usepackage[utf8]{inputenc}
\usepackage{graphicx}
\usepackage{longtable}
\usepackage{booktabs}
\usepackage{csvsimple}
\usepackage{color}
\usepackage{algorithm}
\usepackage{algpseudocode}
\usepackage[export]{adjustbox}
\usepackage{amsmath}
\usepackage{xcolor}
\usepackage{hyperref}
\usepackage{authblk}
\usepackage{pdfpages}
\usepackage{subcaption,comment}
\usepackage{geometry}
\usepackage[export]{adjustbox}
\usepackage{comment}
\usepackage{caption}

\usepackage[style=apa, backend=biber, sorting=nyt]{biblatex}
\usepackage{multicol}
\usepackage{wrapfig}
\definecolor{airforceblue}{rgb}{0.36, 0.54, 0.66}

\newcommand\rev[1]{{\color{black}#1}}

\newcommand\rew[1]{{\color{black}#1}} 

\title{\rew{Compact} 15-minute cities exhibit lower carbon intensity in urban transport}

\author[1,2,3]{Francesco Marzolla}
\author[2,3]{Matteo Bruno}
\author[2,3,4,5]{Hygor P. M. Melo}
\author[1,2,3,6]{Vittorio Loreto}
\affil[1]{Sapienza Univ. of Rome, Physics Dept, Piazzale A. Moro, 5, 00185, Rome, Italy}
\affil[2]{Sony Computer Science Laboratories - Rome, Joint Initiative CREF-SONY, Centro Ricerche Enrico Fermi, Via Panisperna 89/A, 00184, Rome, Italy}
\affil[3]{Centro Ricerche Enrico Fermi (CREF), Via Panisperna 89/A, 00184, Rome, Italy}
\affil[4]{Postgraduate Program in Applied Informatics, University of Fortaleza, 60811-905, Fortaleza, CE, Brazil}
\affil[5]{Núcleo de Ciência de Dados e Inteligência Artificial (NCDIA), University of Fortaleza, 60811-905, Fortaleza, CE, Brazil}
\affil[6]{Complexity Science Hub, Josefst\"{a}dter Strasse 39, A 1080, Vienna, Austria}

\date{}

\begin{document}

\rev{\maketitle}

\begin{abstract}
The 15-minute city concept, which advocates cities where essential services are accessible within a 15-minute walk or bike ride, has gained significant attention in recent years. However, despite being celebrated for promoting sustainability, large-scale empirical evaluations of the effectiveness of the 15-minute concept in reducing emissions \rew{remain limited}. To address this gap, we investigate whether cities with better walking accessibility \rew{to services}, such as 15-minute cities, are associated with lower transportation emissions. \rew{Analysing 662} cities worldwide, we find that cities with better walking accessibility to services emit less CO$_2$ per capita for transport. \rew{An increase of 10 percentage points in the share of residents living in 15-minute accessible areas is associated with an approximate 5\% reduction in transport-related CO$_2$ emissions per capita.} Moreover, among cities with similar \rew{levels of} accessibility, those covering larger areas \rew{and exhibiting lower population densities} tend to emit more. Our findings highlight the effectiveness of decentralised urban planning, especially the proximity-based 15-minute city, in promoting sustainable mobility. \rew{At the same time,} \rev{our results} also emphasise the need to integrate local accessibility \rew{with urban compactness - both in terms of population density and of urbanised area - to support sustainable mobility}.
\end{abstract}

\section*{Introduction}
Road transport is the largest source of CO$_2$ emissions in the European Union, accounting for around a quarter of total emissions~\parencite{EEAAnnualEuropean_2023}, and we observe a similar situation in the US, where 31\% of CO$_2$ emissions are due to transport~\parencite{underwoodDoesSharingBackfire2018}. In a context in which cities are nowadays held responsible for more than 60\% of global greenhouse gases~\parencite{allam202215}, and the urban population is rising at the worldwide scale~\parencite{WorldUrbanizationProspects}, building less car-dependent cities is, therefore, a goal of significant importance in the pathway towards carbon neutrality~\parencite{itf2023transport}. Moreover, drastically diminishing the number of vehicles circulating in our cities would address other externalities of car-centred mobility: degradation of air quality~\parencite{wallington2022vehicle}, deaths and injuries due to road crashes~\parencite{peden2004world}, social exclusion, landscape degradation~\parencite{te2022identifying}, and it would free up public space otherwise necessary to move and park private cars~\parencite{verkade2024movement}.

Since only about $3\%$ of cars worldwide are powered by electricity~\parencite{IEA_veicoli, IEA_veicoli_elettrici}, CO$_2$ emissions still represent a good marker for studying car usage. Therefore, we will examine CO$_2$ emissions resulting from transport in cities also to quantify car usage.

Various urban planning strategies have been proposed to build less car-centred urban environments. 
Among them, the \emph{15-minute city} has recently attracted increasing attention from both researchers and policy-makers~\parencite{sepehriXminuteCitiesGrowing2025}. 
Carlos Moreno introduced this paradigm~\parencite{moreno2021introducing} as a revamping of the compact city~\parencite{haaland2015challenges}, reframing it as a model that promotes urban environments highly accessible by foot and by bicycle.
In a 15-minute city, every basic need of citizens must be fulfilled within a 15-minute radius of their home, either by foot or bicycle. Such proximity of services aims to enable citizens to walk or cycle to essential amenities, reducing their car dependence. The possibility of a shift towards active mobility is believed to improve environmental sustainability, along with health and social cohesion~\parencite{allam202215_net_zero, allam202215, moreno2021introducing, khavarian202315, allam2022theoretical}. 
For this reason, the 15-minute concept has been promoted as part of the post-pandemic Green and Just Recovery Agenda of the C40 Cities, a global network of cities taking action to confront the climate crisis~\parencite{C40}.

However, distributing services more uniformly across urban areas does not always result in lower greenhouse gas emissions. For example, a case study conducted in Beijing between 2000 and 2009~\parencite{wangChangingUrbanForm2014} observed that transitioning toward a more decentralised urban form led to increased commuting distances and higher car usage, thereby increasing CO$_2$ emissions. Even creating infrastructure to encourage active mobility may be ineffective. A case study in three UK municipalities indicated that newly built infrastructures for walking and cycling boosted physical activity but did not significantly reduce CO$_2$ emissions from motorised transport~\parencite{BRAND2014284}. Consequently, the impact of urban planning strategies on transport-related CO$_2$ emissions remains not fully understood.

Nonetheless, certain key features of the 15-minute city do positively impact emission reduction. In particular, mixed land-use and high density, both in terms of population and Points Of Interest (POIs), 
foster sustainable mobility~\parencite{Guneralp_2020}, particularly when coupled with easy access to rail transport~\parencite{choi2020exploring}. 
Several studies find a significant positive correlation between the population density of cities and sustainable mobility patterns~\parencite{newman1996land,wu2021urban}. Specifically, in dense cities studies report lower CO$_2$ emissions from transport~\parencite{gudipudi2016city,baur2014urban,ribeiroEffectsChangingPopulation2019}, fewer vehicle kilometres travelled~\parencite{stone2007compact}, and reduced fuel consumption~\parencite{newman1989gasoline}.
It is argued that this is the case because, while urban sprawl is associated with longer travel distances and thus with extra fuel and resource consumption, high population density often matches with mixed land use and pedestrian-oriented urban forms~\parencite{ye2015sustainable, burton2000compact, yeh2000need}, which are key ingredients of the 15-minute city. 
Land-use mixing, in particular, shows a negative association with both active and motorised transportation~\parencite{FRANK2010S99}. A recent study also revealed that residents in 15-minute cities typically choose closer destinations~\parencite{abbiasov202415}, in line with the previous observation that shorter journeys are observed in cities with a higher density of Points of Interest (POIs)~\parencite{noulasTaleManyCities2012}. \rew{Recent large-scale evidence at the neighbourhood level suggests that areas where essential services are located closer to residents tend to generate lower per-capita transport-related CO$_2$ emissions ~\parencite{marzollaProximitybasedCitiesEmit2026}.} A case study on the Lisbon Metropolitan Area~\parencite{colacco2025does}, testing the effectiveness of the 15-minute city in promoting sustainable mobility, found that
policies combining proximity and density can increase the share of active mobility and public transport, reducing car travel and CO$_2$ emissions. A case study on Quebec City~\parencite{des2017greenhouse} showed that land-use diversity and residential density significantly lower greenhouse gas emissions for transport and motorisation rates, promoting active mobility. Indeed, shifting toward active mobility is known to effectively lower greenhouse gas emissions: analysing data from seven European cities, it was found that individuals who switched their travel mode from cars to bicycles reduced their life cycle CO$_2$ emissions by an average of 3.2 kg per day~\parencite{brand2021climate}.
Nevertheless, systematic, large-scale, data-driven evaluations of the impact of implementing the 15-minute concept on car dependency \rew{remain limited}.

In this work, we aim to determine whether cities that are more accessible on foot, such as 15-minute cities, have lower transport emissions. \rev{We also seek to explain why denser, more mixed-use urban structures tend to generate lower transport emissions, thereby advancing understanding of the well-established association between urban density and emissions. 
To address these gaps, the present study makes three key contributions.
First, we provide the first global-scale analysis linking walking accessibility, measured through a Proximity Time metric, to per-capita transport emissions, using detailed pedestrian-network and service-location data for hundreds of cities. Second, we place accessibility and city area within a unified framework that disentangles their respective effects, showing that accessibility remains strongly associated with per-capita transport emissions even after accounting for city size. Third, we quantify the potential emission reductions that could be achieved if cities matched the accessibility levels of the best performers, offering evidence-based benchmarks to guide urban and transport policy.
}

\rev{
\rew{\section*{Literature Review}}

The idea of a walking city, or proximity city, was born long ago, and took different names and configurations along the years
. In such an urban planning concept, each neighbourhood should allow its residents to access everyday services within walking distance.
The idea is present in the Garden City paradigm
~\parencite{howard1898} and in Perry's Neighbourhood Unit\parencite{perry1929neighbourhood}. Jane Jacobs’s work, from the 1960s, also advocates for dense, mixed-use, walkable areas~\parencite{jacobs1961}. Proposed in the same decade, also the New Urbanism envisions walkable blocks and streets, with housing and shopping in proximity to one another~\parencite {NewUrbanism2000}. 
After them, many urban planning paradigms advocated for the creation of neighbourhoods where, within walking distance, one could reach shops, parks, schools, workplaces, and, generally speaking, services needed for everyday activities \parencite{krier1977,whyte1980social,d2024liveable,d2013simulating,haaland2015challenges}.
In 2016, Carlos Moreno, following these same ideas, proposed the \emph{15-minute city} \parencite{moreno2021introducing}, which is precisely one highly accessible on foot and by bike. The 15-minute city requires that all basic citizen needs, such as work, groceries, healthcare, and leisure, be fulfilled within a 15-minute radius of their residences, by foot or bicycle.
This proximity is designed to reduce car dependence by enabling citizens to use active mobility for reaching essential amenities~\parencite{allam202215_net_zero}, and case studies on single cities showed that it was effective in promoting sustainable mobility \parencite{des2017greenhouse,colacco2025does}. 

Beyond proximity, the paradigm emphasises density, diversity, and digitalisation. Density refers to population concentration, diversity includes cultural variety and mixed land-use, and digitalisation leverages advanced technology and data to facilitate resident engagement and implementation \parencite{moreno2021introducing}.
Some of these key elements of the 15-minute city, namely density and mixed land-use, are known to be correlated with sustainable mobility outcomes. These evidences suggest the 15-minute city as promising in fostering sustainable mobility.

The model is also expected to foster general liveability, to the extent that its principles are utilised to measure liveability \parencite{jeongHowLiveableAre2025}.
However, a fair implementation of the 15-minute paradigm in cities would have to carefully address risks for social justice.
The focus on walking, cycling and other active modes of travel could exclude elderly and disabled residents, if not coupled with an attention to public transport to overcome their mobility restrictions~\parencite{calafiore202220}. The decentralisation and self-sufficiency of neighbourhoods advocated by the 15-minute city could lead to the formation of isolated enclaves, making the city fail to be a unique connected entity where rich and poor, black and white, young and old live together, and, ultimately, to be a place of opportunity~\parencite{glaeser202115, marchigianiUrbanTransitionReturn2022, casarin2023rethinking,hill2024beyond}. A study on a large-scale dataset of mobile GPS traces in the US found that local, \emph{15-minute}, trips might indeed exacerbate socioeconomic segregation, especially among low-income residents~\parencite{abbiasov202415}.
However, social mixing and inclusion are explicit goals of the 15-minute city, falling under the diversity pillar. These objectives are pursued by bringing opportunities to underserved neighbourhoods and incorporating diverse housing types within the same area, thereby attracting residents with different income levels to live together~\parencite{allam2022theoretical}.
However, even social mixing has to be approached with care, since, especially when imposed top-down or pathologising poverty, it can be harmful, as it may lead to social tension and gentrification~\parencite{casarin2023rethinking}.

The 15-minute city concept has also been criticised for being utopian and Eurocentric, raising concerns about its economic feasibility and its applicability in contexts marked by economic and geographic inequalities \parencite{mahmoudpourCriticalDebates15minute2026a}. Critics also note that physical proximity alone overlooks individual preferences and socio-economic differences that may limit the benefits of such infrastructures for certain groups \parencite{noworol15MinuteCityGeographical2022}.

Even at the price of the risks and difficulties exposed, the principles of the 15-minute city are being implemented around the world, in Europe (Paris’s quarter-hour city, Barcelona’s superblocks or ``Rome at your fingertips"~\parencite{roma15mins_2}
), Asia
(Shanghai’s 20-minute
Town or Singapore’s 45-minute city), Australia
(Melbourne’s 20-minute neighbourhoods~\parencite{SHANNON2019100773}), United States
(Portland’s 20-minute neighbourhoods
and Houston’s walkable places \parencite{pozoukidou202115, khavarian2023garden}), 
Latin America
(in Montevideo 
and 
Quito
~\parencite{allam2024mapping}).

Confronting this effort taking place almost at the global scale, it is of primary importance to assess if the 15-minute city is indeed capable of building a sustainable and fair urban environment.
}

\section*{Methods}

\subsection*{Data}

\paragraph{EDGAR}
The Emission Database for Global Atmospheric Research (EDGAR) \rev{v.8.1~\parencite{EDGAR}}, from the EU Joint Research Centre, contains a gridded estimation of air pollutant emissions worldwide, divided by sectors, from 1970 until 2024. Emissions are expressed in the mass of pollutants emitted per unit of time and area. The spatial resolution of the EDGAR dataset is 0.1° of latitude times 0.1° of longitude. Therefore, \rev{along} latitude the resolution is constant in length, being the spacing between datapoints on that direction 11 km; conversely, \rev{along} longitude \rev{— within the cities considered in this study —} it ranges from 4.6 km at 65.55° north in Oulu, Finland, to 10 km at 21.25° north in Honolulu, US. \rev{We filtered the city sample to include only urban areas large enough to be reliably represented at the EDGAR grid resolution.}

\rev{Emissions for each country and sector are estimated by multiplying country-specific activity and technology mix data with corresponding emission factors and reduction factors for abatement systems, taking into consideration also land use, land-use change and forestry~\parencite{monica2022co2}. Activity estimates are based on fuel combustion data~\parencite{monica2022co2}.
Indeed,} the EDGAR dataset builds on the \rev{\textcite{IEA}} estimates of CO$_2$ emissions from fossil fuel combustion.

\rev{
To standardise the methodology and ensure comparability between countries, emission factors are largely default (Tier 1) rather than adapted to local conditions~\parencite{banjaComparativeAnalysisEDGAR2025}.
Accordingly, EDGAR operates mainly as a Tier 1 bottom-up inventory, incorporating elements of Tier 2 methods~\parencite{jrc/ieaGHGEmissionsAll2024}. In practice, this means that EDGAR primarily employs default emission factors to estimate greenhouse gas emissions, while selectively integrating country-specific information where available~\parencite{janssens2019edgar}.
Despite these limitations, for G20 and EU27 countries, EDGAR’s CO\textsubscript{2} emission estimates show strong agreement~\parencite{banjaComparativeAnalysisEDGAR2025} with the official submissions by Parties to the United Nations Framework Convention on Climate Change (UNFCCC). The latter dataset, however, accounts for national or process-specific characteristics through higher-tier methodologies (Tier 2 and Tier 3)~\parencite{schulte2024}.
}

\rev{661 out of 662 cities under consideration belong to Annex I countries of the \textcite{UNFCCC} classification, the only exception being Seoul. For those countries, where reporting systems are more robust, uncertainty in fossil fuel CO\textsubscript{2} emissions is typically within $\pm 5 \% - \pm 10 \%$, at the country level across sectors \parencite{jonesGriddedFossilCO22021}.

Country sector total emissions are then allocated to the 0.1° by 0.1° grid cells using spatial proxy datasets that incorporate the locations of energy and manufacturing facilities, road networks, shipping routes, human and animal population density and agricultural land use. The allocation is done consistently for all
world countries ensuring comparability among countries~\parencite{crippa2018gridded}.

A case study on cities in California and Texas~\parencite{mcdonaldHighresolutionMappingMotor2014} compared CO\textsubscript{2} emissions from vehicles estimated using fuel sales and traffic count data with the estimations of the EDGAR dataset.
The authors found substantial agreement at larger spatial scales; however, the differences in emission estimates increased when focusing on specific urban areas. EDGAR reports on-road CO\textsubscript{2} emissions that are 40–80\% higher than this study’s estimates for California cities and 20–30\% higher for Texas cities. 
As the national inventories are closely aligned, differences in the ways that motor vehicle emissions are disaggregated from national totals down to state and urban scales must account for most of the discrepancies~\parencite{mcdonaldHighresolutionMappingMotor2014}. 
Similar findings were reported by \textcite{Gately2013}, who compared their estimates of vehicle-related CO\textsubscript{2} emissions for Massachusetts, derived from roadway-level traffic data, with those from EDGAR and found that the latter were, on average, 23\% higher.
Building also on these findings, \textcite{mcdonaldHighresolutionMappingMotor2014} note that EDGAR allocates the majority of on-road vehicle emissions within cities, while assigning very little to outlying areas.
Conversely, a study focusing on the Northeastern US~\parencite{gatelyLargeUncertaintiesUrbanScale2017} reports the opposite pattern, finding that EDGAR assigns higher emissions in rural areas compared to the three other inventories considered in the analysis.

Therefore, EDGAR's main limitations lie in deriving national-level emissions from activity data (due to uncertainties in activity data and emission factors \parencite{solazzoUncertaintiesEmissionsDatabase2021}) and, even more, in using spatial proxies to downscale national emissions data to the $0.1° \times 0.1°$ grid level. Nevertheless, this methodology aligns with those currently used for building emission inventories~\parencite{Gately2013}, and the EDGAR dataset is commonly used~\parencite{WB_emissions, lekaki2024road, perez2020air}.
}

This study focuses on CO$_2$ emissions from the transport sector — specifically road and rail transport — in \rev{2023}, excluding those arising from biomass and biofuel combustion (short-cycle carbon).

\paragraph{City boundaries and filtering}
Boundaries of cities
were acquired from the Organisation for Economic Co-operation and Development (OECD)
shapefiles~\parencite{dijkstra2019eu}, focusing on the defined core city \rev{The geographical boundaries of the cities used in this study follow the definition of urban areas provided by the Organisation for Economic Co-operation and Development (OECD)~\parencite{dijkstra2019eu}.
The OECD provides a comprehensive database outlining the metropolitan borders of major cities across different countries.
In this work, we focus on the core areas of cities, as defined by the OECD. A city’s core area is delineated around the central urban unit. It includes all adjacent areas with a population of at least 50,000 inhabitants and a population density of at least 1,500 inhabitants per square kilometre~\parencite{oecd2019fua}.
The core area thus represents the densely populated parts of an urban area.
}.
In the absence of OECD data, we relied on the \rev{city's core defined by the} Global Human Settlement (GHS) files~\parencite{florczyk2019ghs}. However, this second data source is limited to results presented in the Supplementary Information. 

We excluded cities with an area smaller than 44 km$^2$ from our study, to avoid biases due to edge effects.

\paragraph{Population data} 
Population information was sourced from ~\textcite{worldpop}, offering demographic context for our accessibility assessment. We employed a 100m population density grid, refined to correspond with municipal UN population estimates~\parencite{bondarenko2020census}.

\subsection*{Methodology}
\paragraph{\rew{Emissions estimation}}\rev{We computed per-capita CO$_2$ emissions related to road and rail transports for all cities in the sample, \rew{by intersecting the grid provided by the EDGAR dataset with the city boundaries, and multiplying for the appropriate intersection area and 365-day timespan the values of emissions per unit time per unit area of each of the intersected rectangles of the emission grid. We denote per-capita transport-related carbon emissions of a city as $C_\text{pc}$.}

\paragraph{Proximity time $s$}
The proximity of services to home locations in cities has been quantified in various works \parencite{Bruno_et_al_Nature_Cities_2024, nicolettiDisadvantagedCommunitiesHave2023, olivariAreItalianCities2023, vale2023accessibility}. Here we quantify it with the Proximity Time computed by \textcite{Bruno_et_al_Nature_Cities_2024}. 
It consists of the average minutes one citizen needs to walk to reach one of the 20 closest POIs where they can fulfil one of the following needs: learning, healthcare, eating, supplies, moving (with public transport), cultural activities, physical exercise, and other services. 
In the original formulation of the 15-minute city~\parencite{moreno2021introducing}, this time should be no more than 15 minutes for every citizen, for every need. The 15-minute platform developed by~\textcite{15mins} enables users to explore the Proximity Time of different areas in various cities worldwide by showing the metric $s$ on a hexagonal grid composed of elements with a lateral size of $200$ m.

\rev{In detail, the Proximity Time of a city was computed as follows.} \rew{For each hexagon of a grid superimposed on the city}, we identified the 20 nearest POIs for each service category using Open Street Routing Machine (OSRM)~\parencite{luxen2011osrm} for walking routes, \rev{and OSM data from 2023 for POI locations and categories.}}
The Proximity Time for each hexagon is then the average of the scores of each category of services. 
\rev{For further details, we refer to ~\textcite{Bruno_et_al_Nature_Cities_2024}.}

\begin{figure*}[!htb]{}
    \includegraphics[width= \textwidth]{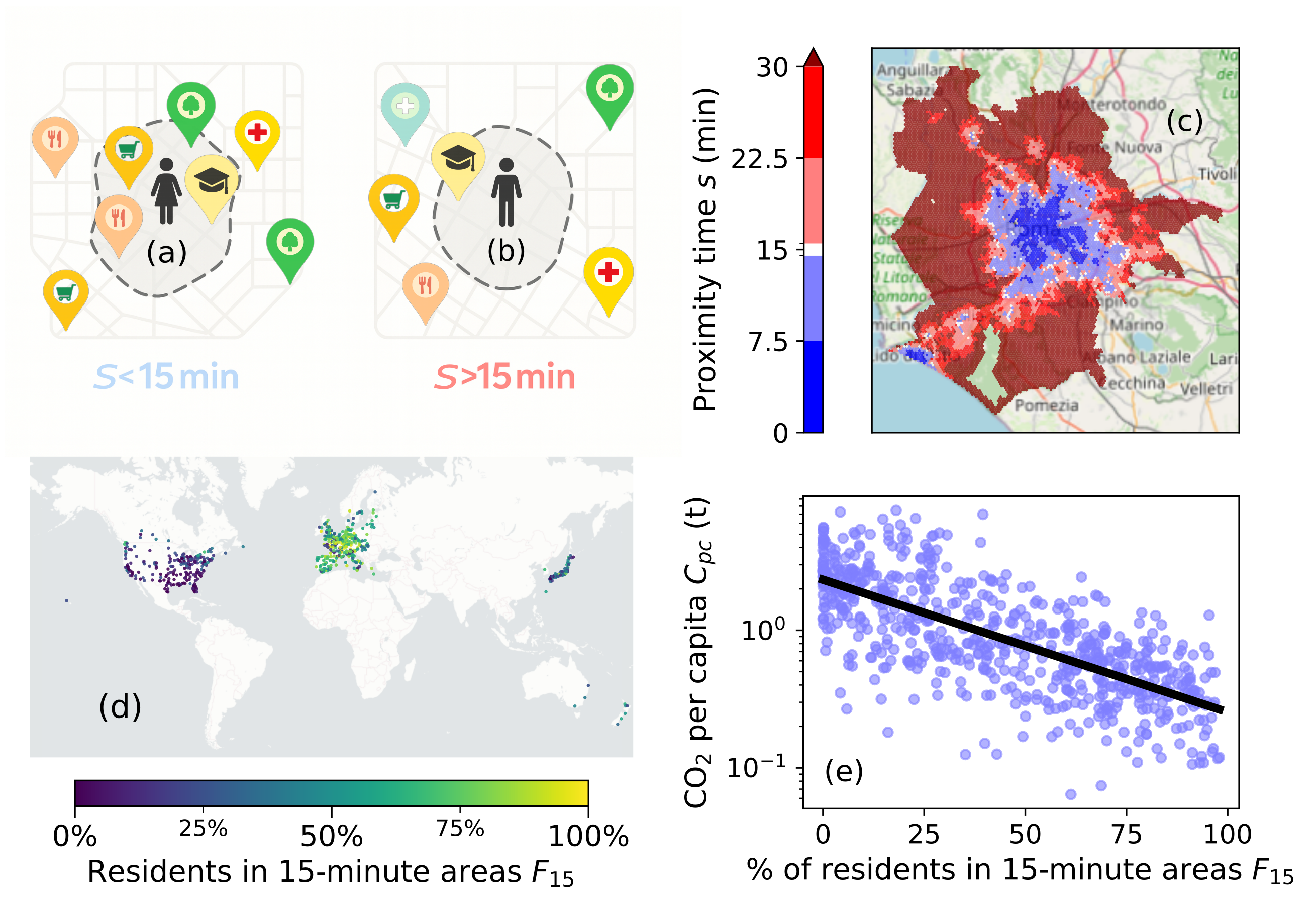}
    \caption{\rev{\textbf{Impact of the fraction of population residing in 15-minute areas of a city on its transport-related CO$_2$ emissions.} Alice, in panel (a), lives in an area with Proximity Time $s < 15$ min: she has to walk \emph{on average} less than 15 minutes to access services. Bob, on the other hand, depicted in panel (b), lives in an area with Proximity Time $s > 15$ min: he has to walk \emph{on average} more than 15 minutes to access services. Pins on the map represent the locations of POIs of different categories (health, education, grocery, restaurants, parks), while the dotted line represents the boundary of the area that is reachable on foot in 15 minutes. On panel (c), the division of Rome into ``15-minute" areas with $s < 15$ min, in shades of blue, and ``non-15-minute" areas with $s > 15$ min, in shades of red. On panel (d), we show a map of the cities considered in the study, with colour encoding the fraction of citizens residing in 15-minute areas $F_{15}$.
    On panel (e), a scatter plot of CO$_2$ emissions per capita versus $F_{15}$, for different cities worldwide. Cities with a higher fraction of the population with access to local services emit less. The black line represents an exponential fit. Copyright of the underlying map of panels (c) and (d) from OpenStreetMap contributors}
        .} 
        \label{fig:co2share15}
\end{figure*}

\rew{Therefore, Proximity Time quantifies, on average, the number of minutes a citizen must walk from their residence to fulfil everyday needs in the city. 
\rev{Fig. (\ref{fig:co2share15}a–b) illustrates Proximity Time through contrasting examples of locations with low and high values.} The lower the value of this metric is, the fewer minutes one needs to walk to access services, and therefore, the better the walking accessibility of the area.}

\paragraph{\rew{Fraction of urban population in 15-minute areas and its relationship with emissions}}Proximity Time was calculated on a fine-grained spatial grid to enable flexible aggregation. \rew{Such a grid consists of hexagons with a side length of 200 m.
The population residing inside each hexagon is derived from the gridded population data provided by WorldPop.
The latter are defined on a square grid with a side of 100 m. We assume the population within each square is uniformly distributed. The population of a hexagon $h$ is calculated as the sum of the population counts of the intersections between $h$ and the squares.

Using the hexagonal grid}, we then derived the share of residents living in 15-minute areas (grid cells with Proximity Time $s < 15$ min).
\rew{We therefore quantify, for each city in our dataset, the fraction of people residing in 15-minute areas, $F_{15}:= P_{15}/P$, where $P_{15}$ indicates the population residing in 15-minute areas, and $P$ the total population of the city~\parencite{logan2022x}.

We fit the emissions for transport per capita, $C_\text{pc}$, and the fraction of people residing in 15-minute areas, $F_{15}$, with an exponential function of the form:
\begin{equation}
    C_\text{pc} = C_0 \, e^{-\frac{ F_{15}}{F_{15, 0}}} \, .
    \label{exp}
\end{equation}}
\noindent \rew{This modelling provides a first estimate of the general trend linking proximity-based accessibility and transport-related emissions across cities. In the following analysis, we build on this trend to examine how other interrelated urban characteristics, in particular urbanised area and population density, influence deviations from it.}
The fit has been performed as a linear least squares regression between the fraction of residents in 15-minute areas $F_{15}$ and the logarithm of emissions $C_\text{pc}$. The confidence intervals on two parameters $C_0$ and $F_{15, 0}$ have been estimated via bootstrapping \parencite{nevitt2001performance}. Running the algorithm for $40 \, 000$ iterations gave the same number of estimations of log$C_0$ and $1/F_{15, 0}$. We deduced an estimation of $C_0$ and $F_{15, 0}$ from each of those estimations. The results are collected in the histograms in SI. The resulting probability density distributions are not Gaussian at 95\% C.L. (K.S. test). Confidence intervals for the fitted parameters have been estimated as symmetric intervals around the mean value enclosing 68$\%$ of the probability.


\paragraph{Z-scores calculation}

\rev{A Z-score, interpreted here as the normalised deviation from the fitted curve, quantifies how much a city's observed value departs from that predicted by its average walking accessibility (Proximity Time). 
\rew{Three distinct Z-scores are computed in this work: (i) for transport-related CO$_2$ emissions, with respect to the exponential scaling introduced in Eq.~\ref{exp}; (ii) for urban area, with respect to a non-parametric kernel regression; and (iii) for urban density, also with respect also to a non-parametric kernel regression.}
\rew{The kernel non-parametric regressions estimate respectively the average surface extension and the average density of cities with a given Proximity Time.} This has been computed using a local linear estimator and a tricube kernel of bandwidth selected by least-squares cross-validation in log-space.
For emissions Z-scores, positive values indicate higher-than-predicted emissions under Eq.~\ref{exp}, whereas negative values indicate lower-than-predicted emissions.
For the area, positive values indicate cities larger than expected at a given proximity time, whereas negative values indicate smaller ones.  For density, positive values indicate cities denser than expected at a given proximity time, whereas negative values indicate less-than-expected dense ones. }

Then, we divided the set of cities into four quartiles based on their Proximity Time. The quartiles are visible in SI. For each city $i$, belonging to a quartile $j$, we calculated the residual $\delta_i$ between the logarithm of its actual area $A_i$ and the logarithm of the average area for its Proximity Time estimated by the kernel regression $\Bar{A}(s_i)$:
\rew{
\begin{equation}
    \delta_i^{(A)} = \text{log}_{10}(A_i) - \text{log}_{10}(\Bar{A}(s_i)) \, .
\end{equation}
Analogously, for density $d_i$ and emissions $C_{pc,i}$ we define
\[
\delta^{(d)}_i = \log_{10}(d_i) - \log_{10}(\Bar{d}(s_i)),
\]
\[
\delta^{(C)}_i = \log_{10}(C_{pc,i}) - \log_{10}(\hat{C}_{pc}(s_i)),
\]
where $\Bar{d}(s_i)$ is the kernel-estimated average density at Proximity Time $s_i$, and $\hat{C}_{pc}(s_i)$ is the expected emission level given by Eq.~\ref{exp}.}

We then computed the standard deviation of the residuals \rew{inside each quartile}, obtaining four standard deviations $\sigma_j$, $j=1,...,4$. The Z-score of the area of a city $i$ belonging to quartile $j$ is therefore given by:
\begin{equation}
Z_i = \frac{\delta_i}{{\sigma_j}} \, .
\end{equation}
\rew{The same standardisation is applied to area, density and emission residuals, yielding Z-scores for each quantity: $Z^{(A)}_i$, $Z^{(d)}_i$ and $Z^{(C)}_i$, respectively.}

\rew{The normality of residual distributions within each quartile was assessed using Kolmogorov–Smirnov tests against a Gaussian distribution with empirical mean and standard deviation.}

\rew{Pearson correlation coefficients were computed to quantify the association between emission Z-scores and both area and density Z-scores.}

\rew{We constructed scatter plots with colour encoding Z-scores, to visually inspect the complex relationship between the key quantities considered, which quantify walking accessibility and urban form, in driving urban transport emissions.}

Finally, we integrated the effects of Proximity Time, \rew{density} and area through \rew{a multivariate regression.}

\rew{\paragraph{Emissions as a function of density, area, and Proximity Time}

We estimate a semi-log linear regression model to examine the association between per capita transport-related CO$_2$ emissions and urban form characteristics. The dependent variable is specified in logarithms. The model takes the following form:

\begin{equation}
\log_{10}(C_{\text{pc}, i}) = \beta_0 
+ \beta_1 \log_{10}(d_i)
+ \beta_2 F_{15, i}
+ \beta_3 \log_{10}(A_i)
+ \varepsilon_i
\label{loglinear_model}
\end{equation}
where $C_{\text{pc}, i}$ denotes transport-related CO$_2$ emissions per capita in city $i$, $d_i$ is population density, $A_i$ is total urban area, and $F_{15, i}$ represents the percentage of residents living in 15-minute areas (0-100 scale). 

The semi-log specification reflects the multiplicative structure and positively skewed distributions, approximately lognormal, of emissions, density, and area, while retaining the accessibility variable in levels given its bounded scale, which includes 0, and distribution highly incompatible with lognormality. Coefficients on log-transformed regressors are interpretable as elasticities, whereas the accessibility coefficient represents a semi-elasticity. Heteroskedasticity-robust (HC3) standard errors are reported throughout.

To assess potential multicollinearity among regressors, we computed variance inflation factors (VIF); all values were within acceptable thresholds, indicating that multicollinearity was not severe and that regularised estimators such as ridge regression were not required.}




\section*{Results}
\subsubsection*{15-minute cities emissions}
We measure the accessibility for pedestrians of each of the cities under study with the Proximity Time $s$. In Fig.~\rev{(\ref{fig:co2share15}c)}, the local values of Proximity Time are shown on a map of Rome. 
We can consider an area to be 15-minute if its Proximity Time $s$ is lower than 15 minutes so that citizens living there have to walk less than a quarter of an hour, on average, to fulfil everyday needs.

The fraction of people residing in 15-minute areas, $F_{15}$, for each city in our dataset, \rev{is sketched, on a map, in Fig.~(\ref{fig:co2share15}d).} 
Such fraction correlates with per-capita transport emissions, $C_{\text{pc}}$, as shown in Fig.~(\ref{fig:co2share15}b), together with the exponential fit provided by Eq.~(\ref{exp}), which yields the following parameter estimate
\rev{
\begin{equation*}
    C_0 = (2.3 \pm 0.1) \text{t}  \, , \quad F_{15, 0} = (45 \pm 2)\% \, .
\end{equation*}
}
\noindent This means that, on average, cities without any 15-minute area, i.e., with $F_{15}=0$, are expected to emit a quantity of CO$_2$ equal to $C_0$ per capita for transport in one year. Since $F_{15, 0} \simeq 1/2$, cities having half of their population residing in 15-minute areas lower their emissions by a factor $e \simeq 2.7$ with respect to $C_0$. Finally, a switch from an utterly non-15-minute city to an utterly 15-minute one would result, following this simple model, in a reduction of more than a factor seven ($\simeq e^2 \simeq 7.4$), with respect to $C_0$.

In the SI, we verified that the observed trend is not spuriously driven by public transport efficiency. We also verified that the result remains robust when using an alternative definition of ``residents in 15-minute areas'', defined as those who need to walk less than 15 minutes to reach \emph{each and every} service required in everyday life.

This finding supports the often-claimed sustainability of the 15-minute city: extending the number of residents living in 15-minute areas, on average, significantly decreases transport carbon emissions.

\subsubsection*{Interplay between walking accessibility, \rew{density, } city \rew{area and CO$_2$ emissions}}
\begin{figure*}[htbp] 
    \includegraphics[width = \textwidth]{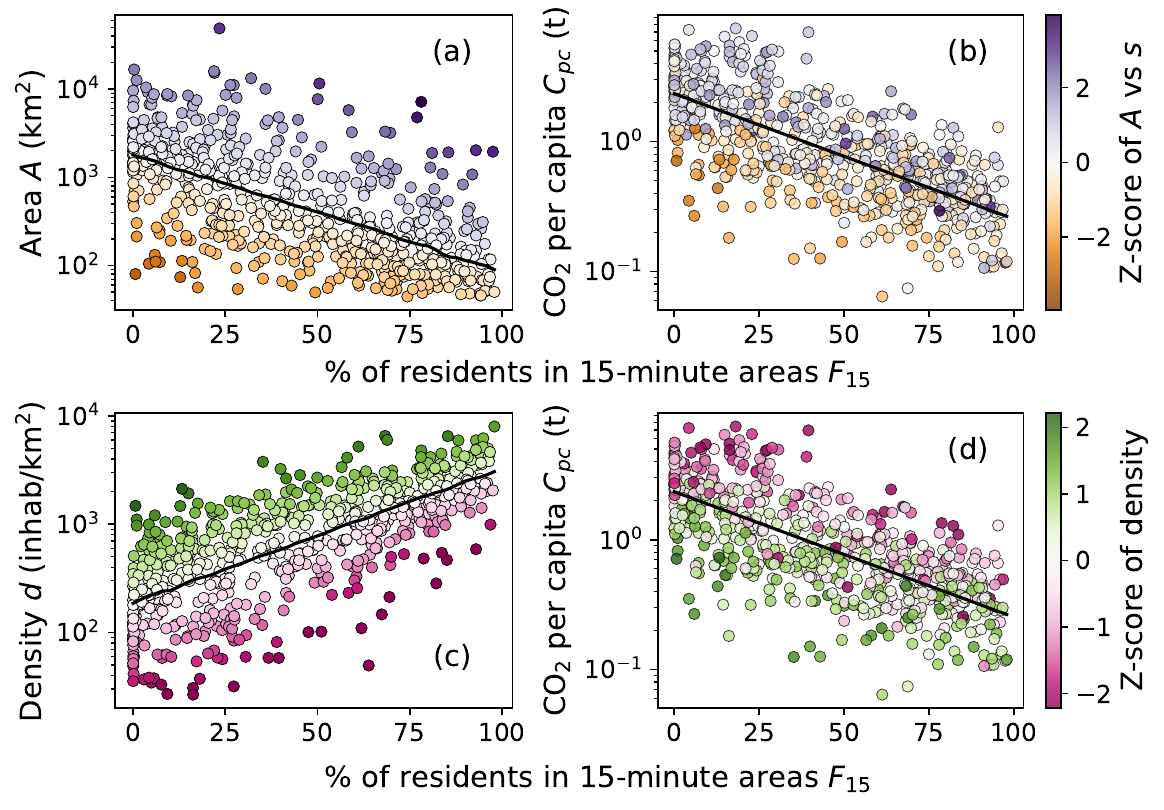}
    \caption{\rew{\textbf{Relationship between fraction of population residing in 15-minute areas ($F_{15}$), urban form, and transport-related CO$_2$ emissions.} 
(a) Urban area $A$ versus $F_{15}$ (log-scale on the vertical axis); the solid line represents a non-parametric kernel regression in log-space. Points are coloured according to the Z-score of the area with respect to the prediction given by the regression. 
(b) Per-capita transport CO$_2$ emissions $C_{pc}$ versus $F_{15}$; the solid line corresponds to the exponential fit defined in Eq.~\ref{exp}. Points are coloured by the same area Z-score of panel (a). 
(c) Urban density $d$ versus $F_{15}$; the solid line shows the kernel regression estimate in log-space. Points are coloured according to the density Z-score. 
(d) Per-capita transport CO$_2$ emissions $C_{pc}$ versus $F_{15}$; the solid line corresponds to Eq.~\ref{exp}. Points are coloured by the same density Z-score of panel (c). 
}}
    \label{fig:Z_scores}
\end{figure*}

\rew{In the previous paragraph, we examined the relationship between accessibility and transport-related carbon emissions.

Moreover, our research reveals the interplay between these quantities and urban form, as characterised by population density and city area.} The relationship between accessibility and city area is evident in 
Fig.~(\ref{fig:Z_scores}a): cities that cover a smaller surface area have a \rew{higher share of residents in 15-minute areas $F_{15}$, indicating} better walking accessibility. We perform a kernel non-parametric regression of city area against \rew{$F_{15}$} to estimate the average trend \rew{($R^2 = 0.41$)}. This result enables us to calculate the Z-score for each city's actual area relative to the area predicted by the regression based on its accessibility. In Fig.~\ref{fig:Z_scores}, we colour the points representing cities according to this Z-score in both panels. Cities with a positive Z-score cover larger areas than the average city at the same accessibility level; conversely, cities with a negative Z-score are smaller than the average city at the same accessibility level. \rew{Fig.~\ref{fig:Z_scores}b shows deviations from the average trend linking higher percentages of residents in 15-minute areas to lower emissions.} These fluctuations can be explained by the Z-score encoded by the colour and, therefore, by the different spatial extent of cities. On average, cities that spread over larger areas than expected, given their walking accessibility, also exhibit higher emissions; conversely, cities occupying smaller areas tend to have lower emissions. This trend can be verified by computing the Z-score of CO$_2$ emissions for cities with respect to the value predicted by the exponential fit (Eq.~\ref{exp}) based on their level of walking accessibility. The correlation coefficient between the two Z-scores described is \rew{0.28 ($p < 0.01$)}.

\rew{We repeat the same procedure using population density $d$ instead of city area. In this case, the relationship between accessibility and density, visible in fig. ~\ref{fig:Z_scores}c, is stronger: the kernel non-parametric regression of density against $F_{15}$ explains a larger fraction of the variance ($R^2 = 0.53$). As before, we compute for each city the Z-score of the observed density relative to the value predicted by the regression given its accessibility level. 
We then compare these density deviations with the deviations of CO$_2$ emissions from the average trend predicted by the exponential fit (Eq.~\ref{exp}). The two quantities are negatively correlated, with a Pearson correlation coefficient of $-0.43$ ($p < 0.01$). This result indicates that cities with higher densities than expected given their level of walking accessibility tend to exhibit lower transport-related emissions, while cities that are less dense than expected tend to show higher emissions. 

Taken together, these findings suggest that both the spatial extent of cities and their population density contribute to explaining deviations from the average accessibility–emissions relationship.
}

Our findings have direct and significant policy implications. Providing proximity-based services alone appears insufficient \rew{to minimise transport-related emissions. Urban densification also contributes to reducing the other two factors that we identify as key drivers of emissions.} This evidence underscores the importance of the 15-minute city concept, which should be complemented by planning and regulatory policies that manage urban expansion and promote more compact urban forms. However, pursuing higher residential density and more diversified land use—both central to achieving proximity of services—may inadvertently exert upward pressure on housing prices~\parencite{des2017greenhouse}.

\rew{
\subsubsection*{Emissions as a function of density, area, and Proximity Time}

\begin{figure*}[htb]
	\centering
    \includegraphics[width=0.7\textwidth]{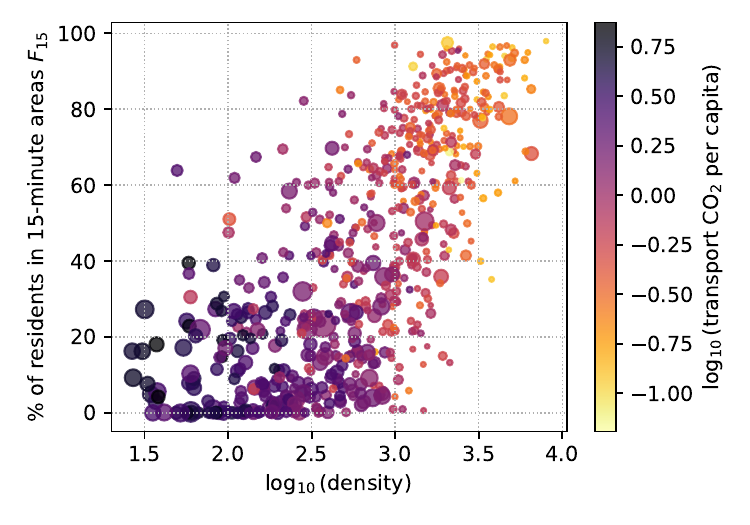} 
        
\caption{\rew{\textbf{Density, Proximity, Area and Emissions interplay.} Distribution of cities in the space defined by the variables used in the model of Eq.~\ref{loglinear_model}. The horizontal axis reports the logarithm of population density ($d$), while the vertical axis shows the share of residents living in 15-minute areas ($F_{15}$, in percent). Colours represent the logarithm of transport-related CO$_2$ emissions per capita ($C_{\text{pc}}$), and the diameter of each marker is proportional to the corresponding city area.}}
\label{fig:modello_finale}
\end{figure*}

Table~\ref{tab:regression} presents the results of the semi-log regression model with heteroskedasticity-robust (HC3) standard errors. The model explains a substantial share of the variation in transport-related CO$_2$ emissions per capita ($R^2 = 0.76$).

Population density is negatively and statistically significantly associated with emissions ($\beta = -0.43 \pm 0.02$, $p < 0.001$). This elasticity implies that a 10\% increase in density is associated with an approximate 4\% reduction in per capita transport emissions, holding other factors constant. 

Urban area exhibits a positive and significant elasticity ($\beta = 0.15 \pm 0.02$, $p < 0.001$), indicating that more spatially extensive cities tend to generate higher transport-related emissions per capita, conditional on density. 

The share of residents living in 15-minute accessible areas is also negatively associated with emissions ($\beta = -0.0023 \pm 0.0004$, $p < 0.001$). Given the base-10 logarithmic specification of the dependent variable, this coefficient implies that an increase of 10 percentage points in the share of residents living in 15-minute accessible areas is associated with an approximate 5\% reduction in transport-related CO$_2$ emissions per capita.

Taken together, the findings suggest that both urban compactness (higher density and low urban area) and proximity-based accessibility are independently associated with lower transport-related CO$_2$ emissions.

\begin{table}[htbp]
\centering
\caption{\rew{Semi-log regression of transport-related CO$_2$ emissions per capita}}
\label{tab:regression}
\begin{tabular}{l c}
\hline
$\log_{10}(\text{density})$ & -0.4341***  (0.0240) \\[4pt]

Accessibility ($F_{15}$, \%) & -0.0023*** (0.0004) \\[4pt]

$\log_{10}(\text{area})$ & 0.1493*** (0.0183) \\[4pt]

Constant & -0.0237 (0.1900) \\

\hline
Observations & 662 \\
$R^2$ & 0.757 \\
Adjusted $R^2$ & 0.756 \\
F-statistic (robust) & 674.0 \\
\hline
\multicolumn{2}{l}{\footnotesize Heteroskedasticity-robust (HC3) standard errors in parentheses.} \\
\multicolumn{2}{l}{\footnotesize *** $p<0.01$, ** $p<0.05$, * $p<0.1$.} \\
\end{tabular}
\end{table}
}

\section*{Discussion}
\rev{This study shows evidence that cities with better walking accessibility tend to have lower per-capita transport emissions, and that this relationship holds even when accounting for city size \rew{and density}. Our comparison of similarly sized \rew{and dense} cities suggests that improving accessibility could yield substantial emission reductions.}

\rev{Our findings contribute to a long-standing debate on how urban spatial structure shapes transport emissions. \textcite{ribeiroEffectsChangingPopulation2019} show that population and area jointly influence citywide CO$_2$ emissions through nonlinear interactions, and that neither density nor population alone sufficiently captures these effects. Their work highlights that the relationship between urban form and emissions depends on city size and structural context, and that density effects are often confounded by correlated variables such as land area. In line with this critique, our results demonstrate that walking accessibility is at least as predictive of transport emissions as density. Although density and proximity \rew{are} correlated, the two are not interchangeable, and proximity offers a more mechanistic interpretation. By explicitly quantifying how shorter walking-access times correspond to lower per-capita transport emissions, we identify a direct behavioural pathway linking urban structure to energy use, complementing previous findings on urban density. This directly addresses a gap identified in a previous review by ~\textcite{hong2021global}, who note that most studies rely on broad proxies such as density or compactness and therefore struggle to isolate mechanisms.}

Building cities that are more accessible on foot, such as the 15-minute city, and fostering active mobility rather than car-dependent mobility, would not only lower carbon emissions but also tackle several other issues caused by cars. One of these is the deterioration of air quality. Taking action in this regard is also imperative, as the WHO warns that air pollution exceeds recommended levels in 83\% of high-income cities and 99\% of low-income cities that monitor their air quality~\parencite{world2023improving}, leading globally to an increased number of patients with respiratory diseases, cancer, and heart diseases~\parencite{allam202215}.
Other drawbacks of cars are \rev{congestion, noise, urban heat effects \parencite{gosslingWhyCitiesNeed2020}}, public space demand, the impact of the maintenance of the road infrastructure and, most importantly, the high rate of \rev{severe injuries and} fatalities due to road \rev{crashes}. Worldwide, the number of people killed in road \rev{crashes} each year is estimated to be almost 1.2 million (15 deaths every 100.000 people \rev{per year) \parencite{GlobalStatusReport2023}}, while the number of people injured could be as high as 50 million~\parencite{peden2004world}. \rev{The burden is disproportionately concentrated in low- and middle-income countries, which account for 92\% of global road-traffic deaths \parencite{GlobalStatusReport2023}.
}

Even though switching city vehicles from internal combustion engines to electric ones would significantly reduce emissions with proper management~\parencite{helmersElectricCarsTechnical2012,panPotentialImpactsElectric2019,arvidssonQuantifyingLifecycleHealth2022}, it would do nothing to address these other issues.
On the contrary, a shift from private vehicles to active mobility and public transport in cities would address all the aforementioned problems.
\rev{
A reduction in the amount of circulating cars in cities would allow for the switch to more sustainable urban forms that would impact the relation with the ecosystem, land-use efficiency, and overall energy demand.

Still, increasing proximity to services and/or density is no easy task. The tension between the space occupied by human activities and the one needed for efficient mobility generates a sub-optimal equilibrium that is hard to escape \parencite{levine2012does}. However, it is also true that cars are the most space-intensive transport mode \parencite{gosslingWhyCitiesNeed2020}. As \textcite{nello-deakinThereSuchThing2019} illustrates, a parked car requires at least three times more space than public transport, and ten times more than the bicycle. When a car is driven at 50 km/ h, it requires 70 times more space than a cyclist or pedestrian.
This means that lowering the number of circulating cars would free up a significant amount of public space that would otherwise be necessary for moving and parking them — space that is arguably not distributed equitably in most cities~\parencite{guzman2021buying}.
Such space could be used for increasing services and to create green and blue corridors and networks, fostering a virtuous cycle of space saving and increased proximity. The ecological solutions would also reduce habitat fragmentation and strengthen ecological resilience, while offering nature-based solutions~\parencite{nesshover2017science} for issues such as the heat island effect and contribute to enhancing well-being and mental health, since both green spaces and active transport are associated with better mental health \parencite{Chekroud2018, Dadvand2016}. Avoiding sprawl and land consumption would preserve natural and agricultural land by concentrating development within a smaller footprint \parencite{ewingCompactnessSprawlReview2015}.
This would also reduce impacts on soil and water systems, while strengthening resilience to heavy rainfall and extreme weather events, which are expected to become more frequent under climate change~\parencite{IPCC2021}.

Despite its global scope, this study has limitations. First, focusing on high-income countries restricts the generalisability of our findings. In many lower-income contexts, OSM data may under-represent services~\parencite{Lagos_slum}, and emission inventories such as EDGAR are less reliable~\parencite{crippa2018gridded}. Structural differences—such as informal transport systems, distinct land-use patterns, and lower vehicle ownership—may also change how proximity translates into emissions~\parencite{UITP, carralesParatransitEmergingTransportation2025, carsmoney}. Additional discussion and analysis for lower-income countries are provided in the SI.

Second, we do not explicitly control for factors such as income or climate. Restricting the sample to high-income countries provides some socio-economic consistency, and we adopt a simple modelling framework to allow comparison across a large number of cities. The SI further shows that differences in public transport availability do not account for the trends observed.

It is important to stress that our analysis does not explicitly model socio-economic and cultural barriers that may prevent residents from benefiting equally from improved walking accessibility. Here, we focus on structural, spatial indicators that can be measured consistently across cities, and we leave a more detailed treatment of these social dimensions to future work and to complementary qualitative and policy-oriented studies.

}

\rev{\section*{Conclusions}
In this work, we conducted a large-scale comparison of cities worldwide and showed that accessibility to proximity services and city \rew{compactness} are two key features shaping the transition from carbon-intensive mobility toward more sustainable patterns.} By analysing more than 600 cities in high-income countries, we showed that, in these regions, cities with greater walking accessibility to essential services significantly reduce CO$_2$ emissions for transport. \rew{An increase of 10 percentage points in the share of residents
living in 15-minute areas is associated with an approximate 5\% reduction in transport-related
CO$_2$ emissions per capita.} As shown in the SI, walking and cycling accessibilities are highly correlated; therefore, the results can be extended to active mobility in general. We can interpret the reduction of CO$_2$ emissions in cities with lower Proximity Time 
as resulting from decreased reliance on private cars, due to the lower car dependency of citizens obtained through enhanced walking accessibility. A shining example of a city accessible by walking is the 15-minute city; our research revealed that cities with a more significant proportion of their population residing in 15-minute areas also exhibit lower transport emissions. Implementing the 15-minute city policy can be a powerful tool in the fight against greenhouse gas emissions and car dependency, offering a promising path towards more sustainable mobility.

A second important outcome of this work is identifying the \rew{density and} area covered by cities as key drivers, intertwined with walking accessibility, for urban CO$_2$ emissions. Since \rew{denser} cities spreading over smaller areas have, on average, better walking accessibility, we had to disentangle the effects. Once we decoupled the effects on CO$_2$ emissions of walking accessibility, city area \rew{and density}, we showed that the fluctuations in CO$_2$ emissions, on top of the general trend linking better accessibility to lower emissions, are well described by fluctuations in area \rew{and density}; cities extending on smaller areas and \rew{having higher population densities} tend to emit less.

The combined influence of walking accessibility, \rew{density and} surface area in explaining CO$_2$ transport emissions of cities has allowed us to develop a model that incorporates these three factors simultaneously. \rew{Its} outcome underscores the significance of our findings and their potential to inform and guide future urban planning and policy decisions: \rev{it is the combination of walking accessibility to proximity services and a \rew{compact}, less sprawling urban area that fosters sustainable mobility in cities.}

In summary, our findings provide large-scale empirical evidence that proximity-based urban forms, such as 15-minute cities, are associated with more sustainable mobility patterns. Moreover, when combined with \rew{compact urban development, the joint effect} of these two features on reducing transport-related emissions becomes even more pronounced. 

To advance sustainability goals, proximity of services must therefore be combined with efficient public transport and urban planning strategies that \rew{densify and} contain urban sprawl.

\section*{Acknowledgements}
Hygor P. M. Melo acknowledges the support of Fundação Edson Queiroz, Universidade de Fortaleza, and Fundação Cearense de Apoio ao Desenvolvimento Científico e Tecnológico.

\section*{Authors' contributions}
Francesco Marzolla: Writing – review \& editing, Writing – original draft, Methodology, Formal analysis, Data curation, Conceptualization. Matteo Bruno: Writing – review \& editing, Methodology, Conceptualization, Supervision. Hygor P.M. Melo: Writing – review \& editing, Methodology, Conceptualization, Supervision. Vittorio Loreto: Writing – review \& editing, Supervision, Funding acquisition.

\section*{Competing interests}
The authors declare no competing interests.

\rev{
\section*{Glossary}
\paragraph{POI} A Point Of Interest is a meaningful geographic location — such as a park, a café, a landmark, a school — historically marked on a map and now represented in digital systems as a point-based datum. Contemporary digital POIs are proxies for real-world destinations, which can support activities like jobs, tourism, recreation, and wayfinding \parencite{psyllidisPointsInterestPOI2022}. POIs could be intended as locations with meaning \parencite{tuanSpacePlacePerspective2011}, therefore they can also be referred to and considered to be \emph{places} \parencite{psyllidisPointsInterestPOI2022}.
}

\section*{Materials and Correspondence}
\paragraph{Supplementary information}
This work contains supplementary material.

\paragraph{Correspondence}
Correspondence and material requests should be addressed to Francesco Marzolla \\ (francesco.marzolla@uniroma1.it).

\rew{\section*{Declaration of generative AI and AI-assisted technologies in the manuscript preparation process}

During the preparation of this work the authors used ChatGPT and Grammarly in order to improve the English language and readability. After using these tools, the authors reviewed and edited the content as needed and take full responsibility for the content of the published article.}

\printbibliography


\newpage 
\appendix



\section*{Supplementary material}

# Supplementary Information

## A  Lower-income countries

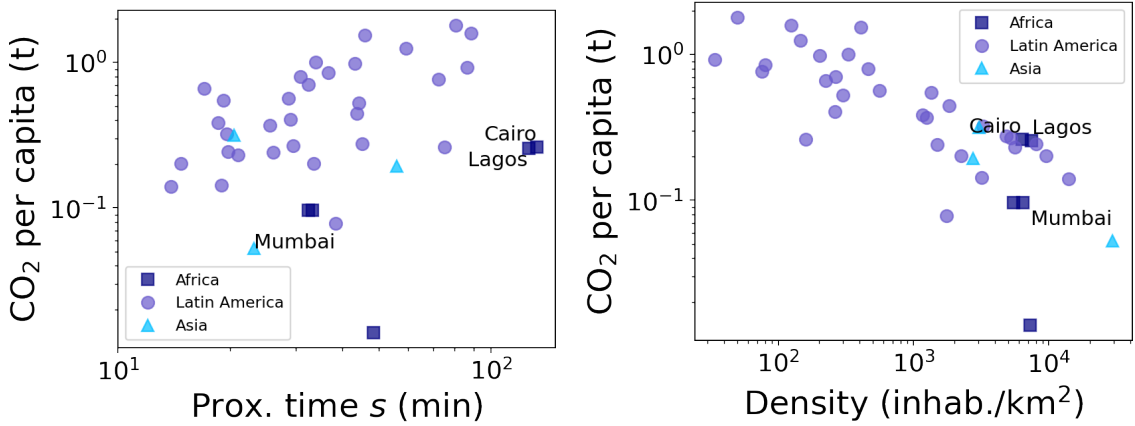

Figure 4: $CO_2$ emissions for transports (road and rail transport) per capita in 2021, in tonnes, versus the average proximity time, in mins, to POIs for city residents in lower-income countries. Each dot represents an urban area, and its dimensions are proportional to the resident population. Different colours and shapes encode different macroregions.

For cities of high-income countries [27], the logarithm of the proximity time $s$ has a linear correlation coefficient of 0.68 $R^2 = 0.68$ with the logarithm of the $CO_2$ for transports per capita. In contrast, cities in lower-income countries display a correlation coefficient of just 0.25.

We hypothesize two factors as possible drivers of the vanishing correlation in the Global South: the lower coverage and reliability of the data used to compute the proximity time $s$ and the higher difficulty for people residing in those areas to own a car.

Regarding the first issue, the destinations of the trips to access everyday services, whose average length constitutes our proximity time $s$, are Points of Interest taken from OpenStreetMap (OSM). OSM is an open platform where private contributors can map the geography of places where they reside, have travelled or even have studied remotely with satellite images. As demonstrated in [46], there is an observed decline in the quality of POIs coverage on OSM in regions with lower socioeconomic status. In particular, Lagos, in Nigeria, is one of the cities that deviate the most from the general approximate power-law relation between proximity and emissions, which holds in high-income countries. In [47], it was shown the existence in 2021 of a slum in Lagos completely not mapped in OSM. The existence of this kind of urban area in developing countries makes it even more challenging to map cities accurately because of the tendency of these areas to be destroyed and rebuilt more rapidly than other types of settlements. To highlight possible overestimations of the proximity time $s$, we can exploit the fact that in the very first approximation, the degree of pedestrian accessibility of services of a city can be estimated by its population density. This is a brute approximation since it does not consider the details of the city's design, whose relation with $CO_2$ emissions is precisely the target of this study. Still, it can give insight into which cities have an estimated accessibility very different from the one expected by their density. Cities that present carbon emissions out of range with respect to their proximity time $s$, such as Mumbai, Cairo and Lagos, align to the general trend if one considers emissions as a function of their density instead. This evidence could be a sign of overestimation of the proximity time for these cities due to a lack of OSM data.

From the side of the different access to cars by people residing in cities located in different areas of the World, in the year 2019 the average vehicle ownership in the EU (motorization rate) was 57 cars per 100 inhabitants [48], in 2021 in the United States it was equal to 84 cars per 100 inhabitants [49], while in 2019 in India it was equal to 23 cars per 100 inhabitants [50] and in 2018 in Nigeria to only six cars per 100 inhabitants [51]. These data can be read as part of a general trend, studied and modelled in [28], which links higher GDP per capita to higher vehicles/population ratio for different countries.

Figure 4b shows the relation between $CO_2$ emissions for transports per capita in 2021 and population density for the cities of lower-income countries.

Because of this looser bond between proximity time and emissions for lower-income countries, we decided to restrict our study to cities in high-income countries, where the dynamics at play are more akin. Fig. 5

shows how the geographical subdivision, among high-income countries, of the cities studied influences their position in the $CO_2$ emissions vs proximity time plane. In particular, cities in the USA and Canada tend to emit more for transport and have worse proximity times with cities in other high-income countries.

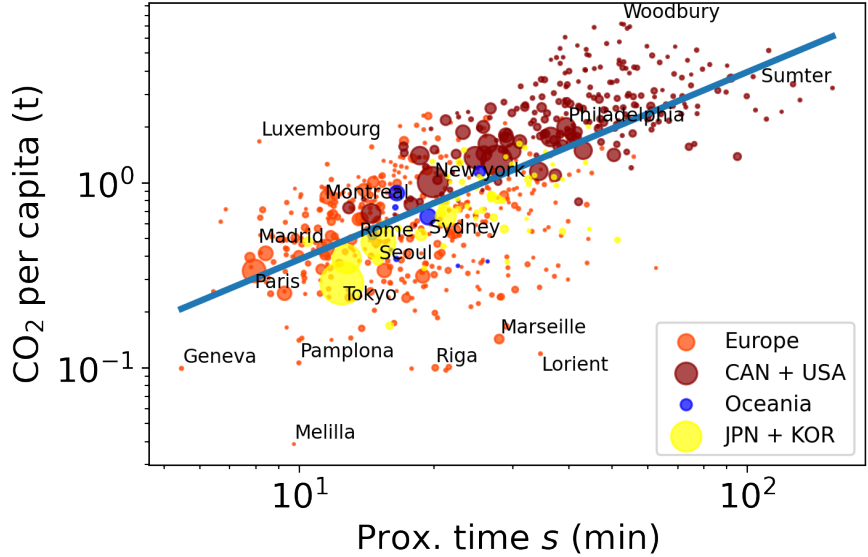

Figure 5: **Geographical subdivision of the studied cities.** $CO_2$ emissions for transports (road and rail transport) per capita in 2021, in tonnes, versus the average proximity time, in mins, of the cities studied. Each dot represents an urban area, and its dimensions are proportional to the resident population. All cities belong to high-income countries, and the macro-region in which they are located is encoded by colour in the plot. CAN stands for Canada, JPN for Japan and KOR for South Corea. The line overimposed on data is the graph of a fitted power law of the form $C_{pc} \sim s^\gamma$, with $\gamma = 1.01 \pm 0.04$.

## B  Correlation between fluctuations on area and on emissions

Fig. (6) depicts, for the cities under study, the relation between fluctuations on area and on emissions respect to the expected values based on the walking accessibility of cities. Fluctuations in emissions are positively correlated with fluctuations in area, with linear correlation coefficient $r = 0.50$. This means that cities covering a larger area than the average for their degree of walking accessibility, tend to emit more for transport, than expected from their degree of walking accessibility.

Fluctuations are quantified via Z-scores in logspace with the following procedure. The Z-score of a city $j$ with area $A_j$ and proximity time $s_j$, belonging to the quartile $i$ of the proximity time distribution, is given by:

$$Z_j = \frac{A_j - \bar{A}(s_j)}{\sigma_i}, \tag{7}$$

where $\bar{A}(s)$ is the average value of area for proximity time $s$, estimated by the kernel regression. The $p$-value at which it is possible to reject the null hypothesis that the residuals at the numerator of Eq. (7) are normally distributed is 0.006 (K.S. test). Being not possible to reject the null hypothesis of normality at $3\sigma$ C.L., we went on computing the standard deviation of residuals in each quartile, $\sigma_i$, and finally computed $Z_j$ for each city. Quartiles are depicted in different colors in Fig. (7a). The same procedure has been used to compute Z-scores of actual $CO_2$ emissions of cities respect to the ones predicted on the basis of their average proximity time. Quartiles and the power-law model used to predict expected emissions are visible in figure (7b).

## C  Confidence intervals on the fitted parameters

The confidence intervals of the parameters of the exponential fit of $C_{pc}$ vs $F_{15}$ have been estimated by bootstrapping. Running the algorithm for 40 000 iterations gave the same number of estimations of $\log A$ and $1/F_{15,0}$. From each of those estimations, we deduced an estimation of $A$ and $F_{15,0}$. The results are collected in the histograms in Fig. (8). The resulting probability density distributions are not Gaussian at 95% C.L..

11

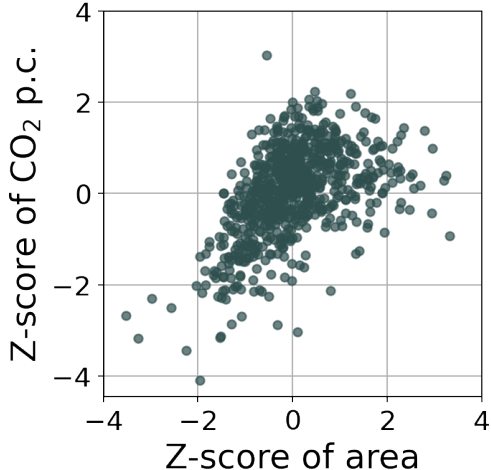

Figure 6: **Fluctuations on city areas are correlated with fluctuations on emissions for transport, with linear correlation coefficient** $r = 0.50$. Each dot represents a city worldwide. On the two axes are depicted respectively the Z-scores of emissions and area respect to some estimation of the expected values of these quantities based on the walking accessibility of the city considered. The expected values are estimated for emissions via a power-law regression of emissions versus proximity time, and for area as a kernel non-parametric regression of area versus proximity time.

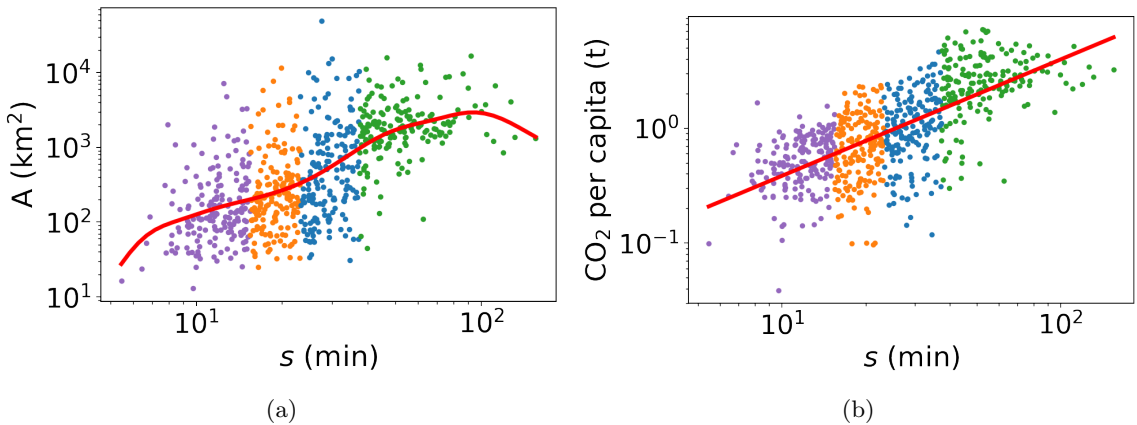

(a)

(b)

Figure 7: **Quartiles, encoded by colour, and fitted models used to estimate Z-scores of city areas (a), and of $CO_2$ emissions (b).**

Confidence intervals for the fitted parameters have been therefore estimated as symmetric intervals around the mean value enclosing 68% of the probability.

## D  Alternative modelling of $CO_2$ emissions

With the same number of parameters as the Cobb-Douglas function used in the main text, we could also model the relationship between emissions, proximity time and city area in the following way, therefore adding an interaction term between proximity time and city area:

$$\log(C_{pc}) = \zeta \log(s) + \xi \log(A) + \phi \log(s) \log(A) + \nu, \qquad (8)$$

which yields $R^2 = 0.63$, therefore an accuracy compatible with the Cobb-Douglas model, which does not include the interaction term. The values of the parameters resulting from a least-squares fit are:

$$\zeta = -1.8 \pm 0.6, \quad \xi = -0.1 \pm 0.1, \quad \phi = 0.26 \pm 0.07, \quad \nu = -0.2 \pm 0.9. \qquad (9)$$

All confidence intervals have been estimated via bootstrapping.

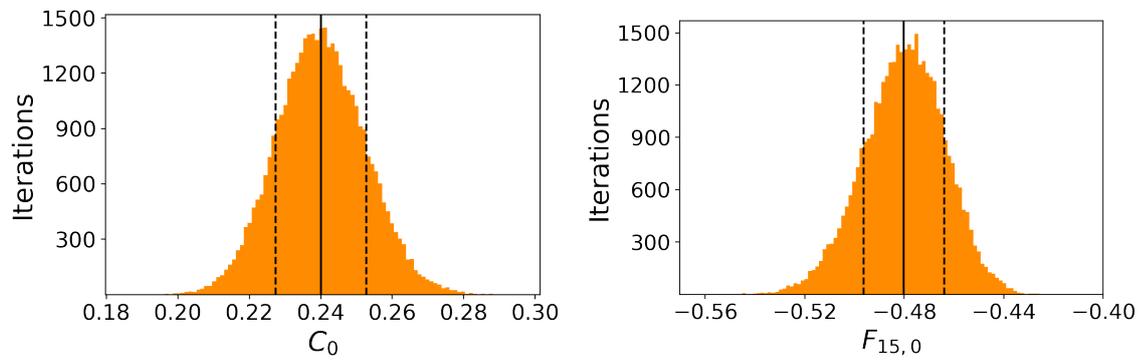

Figure 8: **Probability density functions of the parameters of the exponential fit of $C_{\mathbf{pc}}$ vs $F_{15}$.** The solid line marks the mean value of each distribution, while the dashed lines delimit the confidence interval, corresponding to 68% of the probability. The histograms collect the parameters obtained running for 40 000 iterations of a bootstrapping algorithm.

# E  Cities' metrics

All cities studied are collected, with their key metrics, in table 1.

Table 1: Cities' metrics: population, area, proximity time, and annual $CO_2$ emissions for transport per capita

| City | Country | Population | A (km$^2$) | $s$ (min) | $C_{pc}$ (t) |
|---|---|---|---|---|---|
| Aachen | DEU | 238453 | 163 | 9.4 | 0.78 |
| Aalborg | DNK | 180175 | 1127 | 47.4 | 1.53 |
| Abbotsford | CAN | 154993 | 382 | 19.7 | 1.95 |
| Aberdeen | GBR | 232211 | 185 | 11.2 | 0.81 |
| Ada | USA | 487270 | 2552 | 28.8 | 1.49 |
| Akita | JPN | 285456 | 867 | 19.8 | 1.3 |
| Alachua | USA | 240090 | 2343 | 51.2 | 3.54 |
| Albacete | ESP | 164625 | 1121 | 20.4 | 1.47 |
| Albany | USA | 449673 | 1917 | 34.7 | 3.61 |
| Albuquerque | USA | 749581 | 2510 | 29.2 | 1.88 |
| Alessandria | ITA | 85879 | 198 | 30.5 | 1.32 |
| Algeciras | ESP | 210801 | 96 | 13.0 | 0.14 |
| Alicante | ESP | 408254 | 241 | 18.7 | 0.57 |
| Allen | USA | 360427 | 1704 | 45.5 | 2.87 |
| Almeria | ESP | 204829 | 291 | 19.0 | 0.54 |
| Amiens | FRA | 175338 | 56 | 24.2 | 0.21 |
| Amsterdam | NLD | 1519130 | 1019 | 19.3 | 1.17 |
| Ancona | ITA | 90494 | 124 | 28.2 | 0.73 |
| Angers | FRA | 262839 | 74 | 26.9 | 0.25 |
| Angoulême | FRA | 105754 | 194 | 32.9 | 0.71 |
| Annecy | FRA | 140485 | 70 | 18.1 | 0.29 |
| Antwerp | BEL | 527281 | 199 | 30.3 | 0.75 |
| Aomori | JPN | 265822 | 790 | 29.8 | 1.26 |
| Asahikawa | JPN | 323796 | 684 | 22.6 | 0.82 |
| Ashford | GBR | 134237 | 578 | 25.9 | 1.97 |

Table 1: Cities' metrics: population, area, proximity time, and annual $CO_2$ emissions for transport per capita

| City | Country | Population | A (km$^2$) | $s$ (min) | $C_{pc}$ (t) |
|---|---|---|---|---|---|
| Ashikaga | JPN | 185670 | 195 | 24.9 | 0.93 |
| Athens | GRC | 3514199 | 1917 | 15.5 | 0.34 |
| Atlanta | USA | 2995863 | 3582 | 50.4 | 1.41 |
| Atlantic city | USA | 272180 | 1347 | 54.0 | 2.94 |
| Auckland | NZL | 1475194 | 682 | 14.6 | 0.52 |
| Augsburg | DEU | 299388 | 147 | 8.8 | 0.37 |
| Austin | USA | 1902418 | 5356 | 41.3 | 1.68 |
| Avignon | FRA | 177023 | 75 | 29.2 | 0.35 |
| Badajoz | ESP | 137210 | 1256 | 28.7 | 1.41 |
| Barcelona | ESP | 3496615 | 536 | 9.3 | 0.25 |
| Bari | ITA | 343308 | 346 | 32.3 | 0.36 |
| Barletta | ITA | 86561 | 147 | 15.7 | 0.64 |
| Basel | CHE | 161976 | 23 | 6.4 | 0.26 |
| Basingstoke and deane | GBR | 182496 | 632 | 22.7 | 1.7 |
| Bath and north east somerset | GBR | 184138 | 352 | 15.2 | 1.08 |
| Bedford | GBR | 168255 | 476 | 25.0 | 1.19 |
| Belfast | GBR | 341598 | 138 | 13.8 | 0.62 |
| Belfort | FRA | 94696 | 33 | 25.0 | 0.27 |
| Bell | USA | 371600 | 2634 | 65.9 | 2.64 |
| Benton (ar) | USA | 272614 | 2261 | 68.4 | 2.93 |
| Benton (wa) | USA | 203639 | 3317 | 37.5 | 4.74 |
| Bergamo | ITA | 132503 | 40 | 17.9 | 0.24 |
| Berks | USA | 421660 | 2235 | 58.0 | 2.95 |
| Berlin | DEU | 3683056 | 1078 | 8.4 | 0.42 |
| Bern | CHE | 125417 | 52 | 6.7 | 0.61 |
| Besançon | FRA | 175992 | 409 | 25.1 | 1.11 |
| Bielefeld | DEU | 351932 | 258 | 11.8 | 0.77 |

Table 1: Cities' metrics: population, area, proximity time, and annual $CO_2$ emissions for transport per capita

| City | Country | Population | A (km$^2$) | $s$ (min) | $C_{pc}$ (t) |
|---|---|---|---|---|---|
| Bilbao | ESP | 500184 | 84 | 8.6 | 0.25 |
| Blackburn with darwen | GBR | 237965 | 210 | 18.3 | 1.07 |
| Blackpool | GBR | 139764 | 34 | 11.5 | 0.35 |
| Bologna | ITA | 389299 | 139 | 16.0 | 0.35 |
| Bolzano | ITA | 105738 | 52 | 16.0 | 0.43 |
| Bonn | DEU | 383163 | 175 | 8.2 | 0.65 |
| Boras | SWE | 110333 | 964 | 42.0 | 1.16 |
| Bordeaux | FRA | 717975 | 245 | 20.4 | 0.47 |
| Boston | USA | 3554142 | 4814 | 23.2 | 1.87 |
| Boulder | USA | 295402 | 1796 | 24.2 | 2.2 |
| Boulognesurmer | FRA | 113900 | 175 | 27.2 | 1.04 |
| Bourges | FRA | 92663 | 66 | 33.3 | 0.39 |
| Bournemouth | GBR | 360103 | 114 | 13.1 | 0.45 |
| Bracknell forest | GBR | 117386 | 107 | 17.1 | 1.36 |
| Braga | PRT | 177460 | 181 | 23.7 | 0.66 |
| Brantford | CAN | 108581 | 73 | 22.7 | 0.79 |
| Bratislava | SVK | 399346 | 366 | 13.8 | 0.76 |
| Braunschweig-salzgitter wolfsburg | DEU | 479962 | 621 | 15.3 | 1.09 |
| Brazos | USA | 230776 | 1400 | 40.9 | 2.04 |
| Bremen | DEU | 580387 | 320 | 18.0 | 0.59 |
| Bremerhaven | DEU | 104777 | 74 | 13.2 | 0.43 |
| Brescia | ITA | 204664 | 89 | 22.1 | 0.51 |
| Brest | FRA | 204584 | 48 | 21.3 | 0.1 |
| Brevard | USA | 591679 | 2419 | 50.7 | 1.97 |
| Brighton and hove | GBR | 297800 | 82 | 10.3 | 0.3 |
| Brisbane | AU | 2288460 | 3045 | 25.3 | 1.15 |
| Bristol | GBR | 469260 | 110 | 9.8 | 0.48 |

Table 1: Cities' metrics: population, area, proximity time, and annual $CO_2$ emissions for transport per capita

| City | Country | Population | A (km$^2$) | $s$ (min) | $C_{pc}$ (t) |
|---|---|---|---|---|---|
| Brivelagaillarde | FRA | 81277 | 343 | 42.8 | 1.67 |
| Brno | CZE | 391025 | 230 | 10.5 | 0.37 |
| Broome | USA | 168562 | 1835 | 53.0 | 7.15 |
| Brown | USA | 258295 | 1376 | 49.4 | 3.48 |
| Bruges | BEL | 116290 | 134 | 17.2 | 1.07 |
| Brussels | BEL | 97576 | 56 | 10.1 | 0.83 |
| Budapest | HUN | 1789480 | 522 | 11.2 | 0.34 |
| Burgos | ESP | 167369 | 105 | 11.2 | 0.72 |
| Burnley | GBR | 84853 | 110 | 18.9 | 1.13 |
| Butte | USA | 216494 | 3724 | 103.1 | 3.73 |
| Béziers | FRA | 110013 | 249 | 36.4 | 1.39 |
| Caceres | ESP | 85713 | 1732 | 21.8 | 2.11 |
| Caddo | USA | 230706 | 2192 | 79.5 | 4.1 |
| Cadiz | ESP | 93481 | 30 | 11.8 | 0.14 |
| Caen | FRA | 218111 | 48 | 20.4 | 0.19 |
| Cagliari | ITA | 154322 | 66 | 20.5 | 0.35 |
| Calais | FRA | 92263 | 33 | 23.0 | 0.35 |
| Calgary | CAN | 1376109 | 842 | 20.4 | 1.03 |
| Cambridge | GBR | 134920 | 39 | 8.4 | 0.29 |
| Cameron | USA | 407059 | 2333 | 67.3 | 2.6 |
| Cannes | FRA | 85054 | 24 | 16.5 | 0.26 |
| Cardiff | GBR | 387089 | 141 | 14.2 | 0.56 |
| Carlisle | GBR | 114580 | 979 | 28.8 | 2.09 |
| Cartagena | ESP | 217826 | 542 | 32.7 | 1.43 |
| Caserta | ITA | 88085 | 53 | 25.9 | 0.61 |
| Cass | USA | 169209 | 4540 | 45.5 | 6.89 |
| Catania | ITA | 321776 | 181 | 22.7 | 0.52 |

Table 1: Cities' metrics: population, area, proximity time, and annual $CO_2$ emissions for transport per capita

| City | Country | Population | A (km$^2$) | $s$ (min) | $C_{pc}$ (t) |
|---|---|---|---|---|---|
| Catanzaro | ITA | 83067 | 109 | 62.6 | 0.35 |
| Centre | USA | 152681 | 2630 | 54.7 | 6.95 |
| Ceske budejovice | CZE | 88938 | 54 | 11.1 | 0.3 |
| Chalonsursaône | FRA | 104446 | 365 | 33.8 | 1.03 |
| Chambéry | FRA | 124615 | 222 | 23.9 | 0.86 |
| Champaign | USA | 209653 | 2572 | 38.3 | 4.4 |
| Charleroi | BEL | 509839 | 1433 | 26.2 | 2.22 |
| Charleston | USA | 357038 | 2198 | 76.2 | 1.62 |
| Charlotte | USA | 1133706 | 1408 | 37.7 | 1.57 |
| Chartres | FRA | 108984 | 44 | 39.8 | 0.3 |
| Chatham | USA | 276369 | 919 | 51.6 | 2.15 |
| Cheltenham | GBR | 120106 | 45 | 9.3 | 0.26 |
| Chemnitz | DEU | 244951 | 220 | 11.8 | 1.04 |
| Cherbourg | FRA | 80405 | 68 | 24.5 | 0.38 |
| Cheshire west and chester | GBR | 338895 | 908 | 20.4 | 1.73 |
| Chesterfield | GBR | 106863 | 65 | 15.4 | 0.69 |
| Chicago | USA | 8627748 | 10100 | 24.7 | 1.37 |
| Christchurch | NZL | 387695 | 347 | 16.4 | 0.74 |
| Cincinnati | USA | 886535 | 1489 | 29.8 | 2.38 |
| Clermontferrand | FRA | 282519 | 288 | 21.3 | 0.79 |
| Colchester | GBR | 190515 | 334 | 20.3 | 0.83 |
| Collier | USA | 392080 | 2762 | 66.9 | 2.16 |
| Colmar | FRA | 101080 | 66 | 25.7 | 0.38 |
| Cologne | DEU | 1417693 | 566 | 11.9 | 0.67 |
| Columbus | USA | 1216783 | 1404 | 28.6 | 1.78 |
| Comanche | USA | 123905 | 2533 | 90.2 | 4.28 |
| Como | ITA | 97808 | 36 | 25.3 | 0.26 |

Table 1: Cities' metrics: population, area, proximity time, and annual $CO_2$ emissions for transport per capita

| City | Country | Population | A (km$^2$) | $s$ (min) | $C_{pc}$ (t) |
|---|---|---|---|---|---|
| Cordoba | ESP | 301731 | 1027 | 43.8 | 0.76 |
| Coruna (a) | ESP | 205530 | 36 | 10.1 | 0.14 |
| Cottbus | DEU | 102976 | 160 | 16.7 | 1.13 |
| Coventry | GBR | 729392 | 813 | 15.3 | 0.98 |
| Cracow | POL | 757815 | 326 | 15.0 | 0.63 |
| Crawley | GBR | 112267 | 43 | 15.0 | 0.52 |
| Cumberland (me) | USA | 272230 | 2394 | 55.9 | 4.2 |
| Cumberland (nc) | USA | 312068 | 1599 | 71.3 | 2.34 |
| Cuyahoga | USA | 1347032 | 1784 | 35.4 | 1.83 |
| Dacorum | GBR | 151379 | 212 | 16.9 | 1.22 |
| Dallas | USA | 6760664 | 9237 | 36.4 | 1.76 |
| Dane | USA | 527053 | 3166 | 35.3 | 3.59 |
| Darlington | GBR | 113387 | 196 | 21.3 | 1.39 |
| Darmstadt | DEU | 163918 | 122 | 10.1 | 0.86 |
| Dauphin | USA | 273019 | 1397 | 68.7 | 3.86 |
| Davidson | USA | 658462 | 1351 | 47.9 | 2.47 |
| Debrecen | HUN | 256651 | 461 | 20.6 | 0.63 |
| Delaware | USA | 106422 | 1012 | 43.6 | 4.03 |
| Denver | USA | 2807196 | 8616 | 25.8 | 2.0 |
| Derby | GBR | 275405 | 77 | 11.8 | 0.31 |
| Derry | GBR | 156129 | 1212 | 40.1 | 1.1 |
| Dessau | DEU | 98145 | 244 | 16.3 | 1.16 |
| Detroit (greater) | USA | 3562468 | 5228 | 28.9 | 1.8 |
| Dijon | FRA | 244339 | 68 | 17.5 | 0.31 |
| Doncaster | GBR | 317746 | 566 | 39.2 | 1.56 |
| Donostia-san sebastian | ESP | 170845 | 60 | 11.7 | 0.24 |
| Douai | FRA | 150616 | 65 | 37.9 | 0.36 |

Table 1: Cities' metrics: population, area, proximity time, and annual $CO_2$ emissions for transport per capita

| City | Country | Population | A (km$^2$) | $s$ (min) | $C_{pc}$ (t) |
|---|---|---|---|---|---|
| Douglas (ks) | USA | 113634 | 1204 | 36.5 | 4.61 |
| Douglas (ne) | USA | 550147 | 863 | 25.3 | 1.97 |
| Dresden | DEU | 1342488 | 5777 | 17.1 | 1.68 |
| Dublin | IRL | 498648 | 117 | 12.7 | 0.42 |
| Dundee city | GBR | 146964 | 59 | 13.3 | 0.4 |
| Dunedin | NZL | 113805 | 109 | 26.3 | 0.38 |
| Dunkerque | FRA | 191277 | 78 | 25.1 | 0.22 |
| Durham | USA | 289781 | 767 | 33.6 | 2.39 |
| Dusseldorf | DEU | 807055 | 316 | 10.3 | 0.57 |
| East baton rouge | USA | 443877 | 1151 | 53.9 | 1.46 |
| East staffordshire | GBR | 121594 | 389 | 31.1 | 0.86 |
| Eastbourne | GBR | 106586 | 43 | 12.8 | 0.26 |
| Ector | USA | 146792 | 1557 | 46.6 | 2.99 |
| Edinburgh | GBR | 503142 | 262 | 8.3 | 0.71 |
| Edmonton | CAN | 1183322 | 1980 | 20.0 | 1.51 |
| El paso (co) | USA | 692916 | 4801 | 40.2 | 1.78 |
| Erfurt | DEU | 217929 | 269 | 15.0 | 0.91 |
| Erie (ny) | USA | 842385 | 2709 | 39.6 | 2.75 |
| Erie (pa) | USA | 248682 | 2079 | 43.5 | 4.07 |
| Escambia | USA | 280260 | 1812 | 56.6 | 2.61 |
| Exeter | GBR | 124393 | 47 | 10.2 | 0.29 |
| Falkirk | GBR | 166718 | 293 | 19.0 | 1.93 |
| Fayette | USA | 321264 | 729 | 32.9 | 2.09 |
| Ferrara | ITA | 129419 | 403 | 39.8 | 1.74 |
| Firenze | ITA | 380696 | 117 | 18.8 | 0.33 |
| Flagler-daytona beach | USA | 173334 | 1141 | 73.0 | 2.56 |
| Flensburg | DEU | 85365 | 48 | 11.6 | 0.36 |

Table 1: Cities' metrics: population, area, proximity time, and annual $CO_2$ emissions for transport per capita

| City | Country | Population | A (km$^2$) | $s$ (min) | $C_{pc}$ (t) |
|---|---|---|---|---|---|
| Foggia | ITA | 139634 | 492 | 16.8 | 1.38 |
| Forlì | ITA | 110730 | 191 | 33.0 | 1.06 |
| Forsyth | USA | 363154 | 1068 | 54.4 | 3.01 |
| Frankfurt am main | DEU | 966758 | 369 | 11.8 | 0.62 |
| Freiburg im breisgau | DEU | 232896 | 154 | 9.3 | 0.44 |
| Fresno (greater) | USA | 1027641 | 12664 | 60.9 | 2.59 |
| Fréjus | FRA | 80990 | 109 | 29.4 | 0.78 |
| Fuji | JPN | 363913 | 610 | 26.8 | 1.06 |
| Fujieda | JPN | 298611 | 282 | 23.8 | 0.75 |
| Fukui | JPN | 248304 | 485 | 35.5 | 1.46 |
| Fukuoka | JPN | 2389948 | 1066 | 18.7 | 0.52 |
| Fukushima | JPN | 268582 | 582 | 32.5 | 0.98 |
| Gdansk | POL | 690944 | 396 | 17.7 | 0.5 |
| Genesee | USA | 394597 | 1679 | 64.2 | 2.34 |
| Geneva | CHE | 204702 | 16 | 5.5 | 0.1 |
| Genova | ITA | 567802 | 210 | 16.9 | 0.38 |
| Gent | BEL | 261547 | 156 | 20.7 | 1.37 |
| Gera | DEU | 90682 | 151 | 14.7 | 1.07 |
| Giessen | DEU | 80193 | 72 | 9.8 | 0.93 |
| Gijon | ESP | 237581 | 180 | 16.8 | 0.5 |
| Girona | ESP | 109150 | 38 | 9.4 | 0.17 |
| Glasgow | GBR | 1212675 | 1056 | 17.9 | 1.35 |
| Gloucester | GBR | 132185 | 40 | 15.8 | 0.49 |
| Gottingen | DEU | 115882 | 116 | 6.9 | 0.72 |
| Granada | ESP | 237625 | 88 | 9.3 | 0.33 |
| Graz | AUT | 260087 | 127 | 11.8 | 0.68 |
| Great yarmouth | GBR | 103318 | 177 | 22.1 | 0.53 |

Table 1: Cities' metrics: population, area, proximity time, and annual $CO_2$ emissions for transport per capita

| City | Country | Population | A (km$^2$) | $s$ (min) | $C_{pc}$ (t) |
|---|---|---|---|---|---|
| Greene | USA | 281138 | 1749 | 44.6 | 3.27 |
| Greenville | USA | 486236 | 2006 | 37.6 | 2.21 |
| Grenoble | FRA | 401597 | 80 | 16.3 | 0.17 |
| Groningen | NLD | 186411 | 101 | 29.1 | 0.4 |
| Guadalajara | ESP | 104583 | 220 | 21.7 | 1.02 |
| Guelph | CAN | 130851 | 87 | 16.3 | 0.67 |
| Guildford | GBR | 145627 | 270 | 19.1 | 1.64 |
| Guilford | USA | 521155 | 1698 | 51.3 | 2.85 |
| Gyor | HUN | 106015 | 174 | 27.8 | 0.83 |
| Hachinohe | JPN | 208393 | 304 | 21.5 | 1.08 |
| Hakodate | JPN | 253938 | 647 | 29.2 | 0.37 |
| Halifax | CAN | 412150 | 4908 | 44.4 | 2.13 |
| Halle an der saale | DEU | 236843 | 135 | 11.4 | 0.47 |
| Halton | GBR | 132175 | 78 | 14.8 | 0.8 |
| Hamamatsu | JPN | 622424 | 319 | 16.7 | 0.57 |
| Hamburg | DEU | 1826359 | 737 | 11.8 | 0.52 |
| Hamilton | NZL | 165335 | 110 | 16.1 | 0.43 |
| Hamilton (tn) | USA | 322324 | 1470 | 71.6 | 2.93 |
| Hampden | USA | 447569 | 1622 | 39.7 | 3.39 |
| Hanover | DEU | 536332 | 205 | 10.1 | 0.47 |
| Harrison | USA | 172443 | 1503 | 104.4 | 2.46 |
| Hartford | USA | 893744 | 1944 | 40.3 | 2.66 |
| Heidelberg | DEU | 165241 | 109 | 10.4 | 0.56 |
| Heilbronn | DEU | 121168 | 99 | 12.5 | 0.89 |
| Helsingborg | SWE | 140577 | 346 | 21.6 | 0.71 |
| Helsinki | FIN | 946261 | 768 | 17.4 | 0.41 |
| Hidalgo | USA | 968156 | 3631 | 95.1 | 1.38 |

Table 1: Cities' metrics: population, area, proximity time, and annual $CO_2$ emissions for transport per capita

| City | Country | Population | A (km$^2$) | $s$ (min) | $C_{pc}$ (t) |
|---|---|---|---|---|---|
| Hildesheim | DEU | 100151 | 93 | 13.5 | 0.94 |
| Himeji | JPN | 549842 | 560 | 28.4 | 0.8 |
| Hiroshima | JPN | 1317949 | 1406 | 26.9 | 0.85 |
| Hitachi | JPN | 185145 | 230 | 24.8 | 0.64 |
| Honolulu | USA | 946094 | 1046 | 42.0 | 0.79 |
| Houston | USA | 5642449 | 6737 | 43.1 | 1.49 |
| Hradec kralove | CZE | 90785 | 105 | 15.6 | 0.54 |
| Huelva | ESP | 144528 | 132 | 22.0 | 0.49 |
| Indian river | USA | 159120 | 971 | 46.4 | 2.01 |
| Indianapolis | USA | 1299957 | 2082 | 41.7 | 1.76 |
| Ingham | USA | 258437 | 1433 | 33.2 | 2.86 |
| Ingolstadt | DEU | 142824 | 132 | 13.7 | 0.73 |
| Ipswich | GBR | 150217 | 39 | 12.2 | 0.37 |
| Irakleio | GRC | 132823 | 51 | 17.8 | 0.1 |
| Iserlohn | DEU | 93825 | 125 | 16.3 | 1.09 |
| Isesaki | JPN | 363407 | 604 | 25.5 | 1.05 |
| Jackson (mo) | USA | 1441523 | 3216 | 34.8 | 2.55 |
| Jackson (or) | USA | 205302 | 6705 | 48.6 | 5.64 |
| Jacksonville | USA | 910885 | 1948 | 47.9 | 2.11 |
| Jaen | ESP | 105038 | 347 | 26.3 | 0.95 |
| Jefferson (al) | USA | 579666 | 2808 | 62.0 | 3.13 |
| Jefferson (ky) | USA | 758175 | 1027 | 35.1 | 1.83 |
| Jefferson (tx) | USA | 238450 | 1965 | 77.5 | 2.58 |
| Jena | DEU | 120140 | 114 | 10.9 | 0.47 |
| Jerez de la frontera | ESP | 205312 | 967 | 21.7 | 1.28 |
| Johnson | USA | 143558 | 1571 | 38.8 | 4.75 |
| Kagoshima | JPN | 530884 | 524 | 38.7 | 0.54 |

Table 1: Cities' metrics: population, area, proximity time, and annual $CO_2$ emissions for transport per capita

| City | Country | Population | A (km$^2$) | $s$ (min) | $C_{pc}$ (t) |
|---|---|---|---|---|---|
| Kaiserslautern | DEU | 101019 | 139 | 15.4 | 1.31 |
| Kalamazoo | USA | 239460 | 1479 | 48.4 | 2.48 |
| Kanazawa | JPN | 563361 | 563 | 20.9 | 0.68 |
| Kankakee | USA | 115089 | 1742 | 69.1 | 5.42 |
| Karlsruhe | DEU | 323660 | 172 | 10.5 | 0.55 |
| Kassel | DEU | 201372 | 103 | 9.3 | 0.61 |
| Kaunas | LTU | 271993 | 156 | 16.0 | 0.29 |
| Kecskemet | HUN | 103891 | 322 | 26.5 | 0.93 |
| Kent | USA | 592801 | 2252 | 42.2 | 2.54 |
| Kern | USA | 999672 | 16724 | 92.0 | 3.67 |
| Kettering | GBR | 106002 | 230 | 17.7 | 1.0 |
| Kiel | DEU | 255687 | 112 | 10.2 | 0.32 |
| Kingston upon hull | GBR | 267936 | 70 | 17.0 | 0.3 |
| Kitakyushu | JPN | 980261 | 508 | 28.7 | 0.56 |
| Kitchener | CAN | 498106 | 319 | 17.7 | 0.82 |
| Klaipeda | LTU | 125485 | 87 | 17.2 | 0.29 |
| Knox | USA | 454107 | 1361 | 54.2 | 2.24 |
| Koblenz | DEU | 115639 | 105 | 15.7 | 0.88 |
| Kochi | JPN | 307138 | 313 | 21.9 | 0.84 |
| Kofu | JPN | 299627 | 316 | 21.4 | 0.78 |
| Koriyama | JPN | 313218 | 734 | 31.9 | 1.54 |
| Kosice | SVK | 240103 | 242 | 16.8 | 0.34 |
| Krefeld | DEU | 220583 | 136 | 13.2 | 0.8 |
| Kumamoto | JPN | 815454 | 536 | 27.2 | 0.65 |
| Kurume | JPN | 293057 | 231 | 33.5 | 1.09 |
| Kusatsu | JPN | 273752 | 164 | 17.6 | 0.55 |
| Kushiro | JPN | 171969 | 1205 | 44.5 | 0.56 |

Table 1: Cities' metrics: population, area, proximity time, and annual $CO_2$ emissions for transport per capita

| City | Country | Population | A (km$^2$) | $s$ (min) | $C_{pc}$ (t) |
|---|---|---|---|---|---|
| Kyoto | JPN | 1467248 | 828 | 10.5 | 0.48 |
| København | DNK | 1119969 | 392 | 14.9 | 0.47 |
| Lackawanna | USA | 190470 | 1184 | 43.9 | 4.92 |
| Lafayette | USA | 246881 | 691 | 61.8 | 2.06 |
| Lafayette (in) | USA | 187487 | 1282 | 39.4 | 2.36 |
| Lancaster (ne) | USA | 303520 | 2166 | 30.6 | 2.51 |
| Lancaster (pa) | USA | 544923 | 2535 | 55.7 | 2.91 |
| Lane | USA | 352527 | 11131 | 61.4 | 4.86 |
| Larimer | USA | 323162 | 5717 | 37.0 | 3.04 |
| Las cruces | USA | 229947 | 6689 | 111.7 | 5.17 |
| Las vegas | USA | 2649719 | 10470 | 37.3 | 1.1 |
| La Spezia | ITA | 87475 | 56 | 25.7 | 0.37 |
| Latina | ITA | 111889 | 276 | 33.4 | 0.91 |
| Lausanne | CHE | 141237 | 41 | 8.2 | 0.42 |
| Lecce | ITA | 110531 | 236 | 25.6 | 0.62 |
| Lee | USA | 813542 | 2061 | 74.1 | 1.56 |
| Leeds | GBR | 2162520 | 1667 | 14.6 | 1.01 |
| Lehavre | FRA | 235442 | 167 | 23.3 | 0.46 |
| Lehigh | USA | 684246 | 1878 | 43.0 | 2.51 |
| Leicester | GBR | 380886 | 72 | 14.2 | 0.37 |
| Leipzig | DEU | 538550 | 297 | 10.3 | 0.67 |
| Lemans | FRA | 183545 | 178 | 22.3 | 0.74 |
| Lensliévin | FRA | 235844 | 213 | 36.7 | 1.03 |
| Leon | ESP | 116374 | 38 | 13.7 | 0.24 |
| Liberec | CZE | 104717 | 105 | 14.7 | 0.63 |
| Liege | BEL | 803074 | 1966 | 19.4 | 2.29 |
| Lille | FRA | 1063111 | 258 | 20.5 | 0.41 |

Table 1: Cities' metrics: population, area, proximity time, and annual $CO_2$ emissions for transport per capita

| City | Country | Population | A (km$^2$) | $s$ (min) | $C_{pc}$ (t) |
|---|---|---|---|---|---|
| Limoges | FRA | 201700 | 99 | 26.6 | 0.52 |
| Lincoln | GBR | 97906 | 35 | 16.0 | 0.29 |
| Linn | USA | 216945 | 1841 | 39.4 | 3.29 |
| Lisbon | PRT | 1196211 | 1090 | 9.8 | 0.5 |
| Liverpool | GBR | 1259404 | 555 | 15.2 | 0.74 |
| Livorno | ITA | 147395 | 104 | 18.9 | 0.34 |
| Ljubljana | SVN | 302352 | 275 | 14.3 | 0.97 |
| Lleida | ESP | 142737 | 211 | 17.3 | 0.79 |
| Logrono | ESP | 152547 | 77 | 12.2 | 0.5 |
| Lorca | ESP | 96801 | 1662 | 57.3 | 2.24 |
| Lorient | FRA | 171696 | 30 | 34.6 | 0.12 |
| Los angeles (greater) | USA | 18026424 | 48918 | 27.6 | 1.31 |
| Lubbock | USA | 302840 | 2292 | 31.2 | 3.31 |
| Lubeck | DEU | 222577 | 209 | 14.2 | 0.8 |
| Lucas | USA | 408236 | 904 | 33.4 | 2.16 |
| Lugo | ESP | 84129 | 328 | 20.1 | 1.13 |
| Luton | GBR | 219527 | 42 | 14.7 | 0.2 |
| Luxembourg | LUX | 122749 | 51 | 8.1 | 1.67 |
| Luzerne | USA | 281499 | 2268 | 63.8 | 4.46 |
| Lyon | FRA | 1306260 | 219 | 17.3 | 0.24 |
| Madison | USA | 361224 | 1982 | 61.0 | 2.49 |
| Madrid | ESP | 5396933 | 1292 | 11.4 | 0.44 |
| Magdeburg | DEU | 245292 | 199 | 13.1 | 0.73 |
| Mahoning | USA | 206764 | 1090 | 47.0 | 4.21 |
| Maidstone | GBR | 173604 | 395 | 26.0 | 1.79 |
| Mainz | DEU | 231658 | 96 | 9.4 | 0.49 |
| Malaga | ESP | 722306 | 440 | 12.0 | 0.51 |

Table 1: Cities' metrics: population, area, proximity time, and annual $CO_2$ emissions for transport per capita

| City | Country | Population | A (km$^2$) | $s$ (min) | C$_{pc}$ (t) |
|---|---|---|---|---|---|
| Malmo | SWE | 483091 | 600 | 14.5 | 0.4 |
| Manchester | GBR | 2905881 | 1274 | 14.0 | 0.91 |
| Mannheim-ludwigshafen | DEU | 561554 | 308 | 12.2 | 0.65 |
| Mansfield | GBR | 110616 | 76 | 14.7 | 0.62 |
| Marbella | ESP | 212917 | 126 | 19.0 | 0.27 |
| Maribor | SVN | 106238 | 148 | 17.4 | 1.21 |
| Marion (fl) | USA | 381418 | 4065 | 126.6 | 3.4 |
| Marion (or) | USA | 332844 | 2935 | 44.4 | 2.96 |
| Marseille | FRA | 1487478 | 297 | 28.0 | 0.14 |
| Marugame | JPN | 118017 | 120 | 21.8 | 1.0 |
| Matsumoto | JPN | 230324 | 810 | 28.9 | 1.38 |
| Matsuyama | JPN | 569580 | 853 | 26.8 | 0.86 |
| Mclean | USA | 177917 | 3037 | 54.0 | 6.81 |
| Mclennan | USA | 226574 | 2578 | 80.6 | 3.17 |
| Mechelen | BEL | 88865 | 65 | 13.6 | 1.07 |
| Medway | GBR | 275331 | 190 | 18.3 | 0.62 |
| Melbourne | AU | 4095505 | 2840 | 16.5 | 0.88 |
| Melilla | ESP | 82351 | 12 | 9.7 | 0.04 |
| Melun | FRA | 104575 | 82 | 25.9 | 0.72 |
| Memphis | USA | 882756 | 1879 | 60.1 | 2.33 |
| Merced | USA | 292925 | 4690 | 71.5 | 3.69 |
| Mesa | USA | 173061 | 6501 | 44.4 | 4.87 |
| Messina | ITA | 202517 | 211 | 33.1 | 0.55 |
| Metz | FRA | 218985 | 73 | 25.5 | 0.37 |
| Miami (greater) | USA | 5934671 | 8476 | 34.3 | 1.15 |
| Middlesbrough | GBR | 442093 | 346 | 18.9 | 0.85 |
| Midland | USA | 154810 | 2042 | 42.0 | 3.51 |

Table 1: Cities' metrics: population, area, proximity time, and annual $CO_2$ emissions for transport per capita

| City | Country | Population | A (km$^2$) | s (min) | C$_{pc}$ (t) |
|---|---|---|---|---|---|
| Milan | ITA | 4014526 | 1953 | 22.1 | 0.55 |
| Milton keynes | GBR | 293420 | 306 | 14.5 | 0.86 |
| Milwaukee | USA | 918768 | 624 | 17.2 | 1.45 |
| Minneapolis | USA | 2046883 | 3506 | 26.8 | 2.52 |
| Minnehaha | USA | 179500 | 2095 | 45.6 | 4.13 |
| Miskolc | HUN | 126200 | 236 | 19.5 | 0.53 |
| Mito | JPN | 414974 | 315 | 23.1 | 1.01 |
| Miyazaki | JPN | 377500 | 637 | 37.0 | 1.07 |
| Mobile | USA | 375535 | 3040 | 99.8 | 2.93 |
| Modena | ITA | 181577 | 182 | 25.5 | 0.89 |
| Monchengladbach | DEU | 263264 | 169 | 12.1 | 0.78 |
| Monroe (in) | USA | 139892 | 1051 | 35.8 | 1.66 |
| Mons | BEL | 250777 | 538 | 18.3 | 2.04 |
| Monterey | USA | 405046 | 7507 | 52.7 | 4.17 |
| Montgomery (al) | USA | 218200 | 1787 | 68.4 | 2.59 |
| Montgomery (oh) | USA | 485928 | 1201 | 42.1 | 2.59 |
| Montpellier | FRA | 421827 | 67 | 20.1 | 0.2 |
| Montreal | CAN | 3432357 | 1592 | 17.8 | 0.76 |
| Morioka | JPN | 265246 | 849 | 37.5 | 0.96 |
| Muenster | DEU | 342252 | 303 | 11.9 | 0.75 |
| Mulhouse | FRA | 254206 | 75 | 23.8 | 0.32 |
| Munich | DEU | 1609658 | 310 | 7.8 | 0.39 |
| Murcia | ESP | 460143 | 879 | 27.1 | 1.16 |
| Muscogee | USA | 184236 | 531 | 52.6 | 1.8 |
| Muskegon | USA | 157704 | 1349 | 62.1 | 2.9 |
| Nagano | JPN | 362020 | 789 | 23.3 | 1.5 |
| Nagasaki | JPN | 443237 | 444 | 51.4 | 0.49 |

Table 1: Cities' metrics: population, area, proximity time, and annual $CO_2$ emissions for transport per capita

| City | Country | Population | A (km$^2$) | $s$ (min) | $C_{pc}$ (t) |
|---|---|---|---|---|---|
| Nagoya | JPN | 7589542 | 4032 | 21.2 | 0.66 |
| Naha | JPN | 836559 | 202 | 15.9 | 0.17 |
| Namur | BEL | 113914 | 174 | 16.8 | 2.02 |
| Nancy | FRA | 256846 | 67 | 15.5 | 0.29 |
| Nantes | FRA | 589061 | 150 | 20.0 | 0.32 |
| Napa | USA | 141039 | 1587 | 72.5 | 2.66 |
| Napoli | ITA | 3397273 | 1282 | 18.9 | 0.31 |
| Nashville | USA | 356216 | 1599 | 63.5 | 2.14 |
| Neumunster | DEU | 80184 | 71 | 14.6 | 0.58 |
| New hanover | USA | 238045 | 468 | 55.1 | 1.2 |
| New haven | USA | 857343 | 1607 | 34.0 | 2.0 |
| New orleans | USA | 647984 | 1033 | 47.2 | 1.66 |
| New york | USA | 17410965 | 11604 | 19.9 | 1.01 |
| Newcastle upon tyne | GBR | 864948 | 406 | 12.0 | 0.84 |
| Newport | GBR | 153835 | 191 | 16.4 | 1.24 |
| Newport news | USA | 294549 | 303 | 33.9 | 1.4 |
| Niagara falls | CAN | 85123 | 213 | 22.5 | 2.37 |
| Nice | FRA | 660216 | 210 | 30.3 | 0.26 |
| Niigata | JPN | 330937 | 178 | 19.0 | 0.34 |
| Niort | FRA | 102560 | 499 | 38.6 | 2.23 |
| North east lincolnshire | GBR | 161355 | 192 | 22.8 | 0.61 |
| Northampton | GBR | 226438 | 79 | 16.9 | 0.51 |
| Norwich | GBR | 142776 | 39 | 9.7 | 0.4 |
| Nottingham | GBR | 345422 | 74 | 8.7 | 0.51 |
| Novara | ITA | 100910 | 102 | 23.2 | 0.91 |
| Nueces | USA | 339619 | 2027 | 61.2 | 1.93 |
| Numazu | JPN | 457340 | 449 | 21.0 | 0.79 |

Table 1: Cities' metrics: population, area, proximity time, and annual $CO_2$ emissions for transport per capita

| City | Country | Population | A (km$^2$) | $s$ (min) | $C_{pc}$ (t) |
|---|---|---|---|---|---|
| Nuremberg | DEU | 762520 | 327 | 11.0 | 0.56 |
| Nyiregyhaza | HUN | 134398 | 274 | 24.7 | 0.67 |
| Nîmes | FRA | 233749 | 538 | 35.3 | 0.98 |
| Obihiro | JPN | 157427 | 517 | 32.6 | 0.55 |
| Oita | JPN | 549530 | 624 | 25.7 | 0.71 |
| Okayama | JPN | 1137535 | 1144 | 27.7 | 0.81 |
| Oklahoma | USA | 1019132 | 3239 | 46.7 | 2.82 |
| Oldenburg (oldenburg) | DEU | 168483 | 103 | 11.4 | 0.93 |
| Olomouc | CZE | 98699 | 103 | 14.0 | 0.44 |
| Omuta | JPN | 118289 | 79 | 32.2 | 0.64 |
| Onondaga | USA | 430117 | 2061 | 40.0 | 4.0 |
| Orange | USA | 1872129 | 2992 | 41.7 | 1.88 |
| Orebro | SWE | 150377 | 1494 | 31.7 | 1.13 |
| Orléans | FRA | 273639 | 324 | 24.8 | 0.66 |
| Osaka | JPN | 15490091 | 4785 | 12.8 | 0.39 |
| Oslo | NOR | 579375 | 452 | 13.8 | 0.16 |
| Osnabruck | DEU | 156119 | 120 | 11.6 | 0.94 |
| Ostrava | CZE | 390774 | 304 | 21.7 | 0.54 |
| Ottawa | CAN | 1341174 | 3200 | 17.1 | 1.57 |
| Oulu | FIN | 163758 | 2751 | 45.0 | 1.13 |
| Ourense | ESP | 92021 | 84 | 20.9 | 0.78 |
| Outagamie | USA | 184373 | 1630 | 51.8 | 4.59 |
| Oviedo | ESP | 198077 | 186 | 15.1 | 0.76 |
| Oxford | GBR | 169293 | 45 | 8.5 | 0.31 |
| Paderborn | DEU | 153724 | 179 | 11.2 | 0.95 |
| Padova | ITA | 239621 | 112 | 26.1 | 0.61 |
| Palermo | ITA | 637654 | 189 | 28.2 | 0.24 |

Table 1: Cities' metrics: population, area, proximity time, and annual $CO_2$ emissions for transport per capita

| City | Country | Population | A (km$^2$) | $s$ (min) | $C_{pc}$ (t) |
|---|---|---|---|---|---|
| Palma | ESP | 437407 | 119 | 11.1 | 0.25 |
| Palma de mallorca | ESP | 438608 | 204 | 11.7 | 0.35 |
| Pamplona | ESP | 195933 | 24 | 10.0 | 0.11 |
| Panevezys | LTU | 86933 | 49 | 17.4 | 0.25 |
| Pardubice | CZE | 89324 | 82 | 16.9 | 0.49 |
| Paris | FRA | 9864887 | 2027 | 7.9 | 0.33 |
| Parma | ITA | 177636 | 261 | 29.4 | 1.05 |
| Pau | FRA | 149023 | 66 | 19.2 | 0.28 |
| Pecs | HUN | 122444 | 163 | 19.7 | 0.59 |
| Peoria | USA | 175124 | 1608 | 53.7 | 4.14 |
| Perpignan | FRA | 244826 | 79 | 34.4 | 0.27 |
| Perugia | ITA | 163727 | 447 | 39.8 | 0.62 |
| Pesaro | ITA | 89830 | 126 | 32.6 | 0.68 |
| Pescara | ITA | 121248 | 33 | 21.3 | 0.26 |
| Peterborough | CAN | 81161 | 66 | 18.2 | 0.59 |
| Pforzheim | DEU | 119515 | 97 | 12.5 | 0.62 |
| Philadelphia | USA | 6000249 | 11146 | 39.4 | 2.0 |
| Philadelphia (greater) | USA | 4238474 | 4427 | 25.9 | 1.43 |
| Phoenix | USA | 4548560 | 15320 | 30.0 | 1.48 |
| Piacenza | ITA | 100463 | 117 | 23.9 | 1.21 |
| Pima | USA | 1085485 | 15966 | 46.6 | 1.96 |
| Pisa | ITA | 87976 | 182 | 23.9 | 1.06 |
| Pitt | USA | 192260 | 1614 | 77.3 | 2.36 |
| Pittsburgh | USA | 1102286 | 1927 | 31.9 | 2.56 |
| Plymouth | GBR | 267441 | 80 | 11.6 | 0.27 |
| Plzen | CZE | 172290 | 136 | 14.4 | 0.69 |
| Polk | USA | 467947 | 1501 | 28.7 | 2.72 |

Table 1: Cities' metrics: population, area, proximity time, and annual $CO_2$ emissions for transport per capita

| City | Country | Population | A (km$^2$) | $s$ (min) | $C_{pc}$ (t) |
|---|---|---|---|---|---|
| Portland | USA | 1882994 | 4490 | 21.2 | 2.17 |
| Portsmouth | GBR | 306925 | 65 | 10.2 | 0.33 |
| Potter | USA | 252326 | 4195 | 53.0 | 4.48 |
| Prague | CZE | 1416104 | 532 | 11.4 | 0.41 |
| Prato | ITA | 189338 | 96 | 25.1 | 0.53 |
| Preston | GBR | 150547 | 142 | 16.0 | 0.98 |
| Providence | USA | 759269 | 1575 | 35.3 | 2.19 |
| Pueblo | USA | 164084 | 4463 | 54.6 | 3.9 |
| Pulaski | USA | 368090 | 2006 | 56.8 | 4.17 |
| Punta gorda | USA | 170581 | 1375 | 131.2 | 2.79 |
| Quebec | CAN | 586593 | 463 | 19.2 | 1.22 |
| Quimper | FRA | 85256 | 290 | 34.6 | 1.18 |
| Racine | USA | 192951 | 875 | 39.1 | 2.8 |
| Ravenna | ITA | 138740 | 648 | 55.8 | 1.38 |
| Reading | GBR | 160583 | 39 | 10.1 | 0.5 |
| Red deer | CAN | 104464 | 104 | 25.7 | 0.91 |
| Redditch | GBR | 88245 | 54 | 18.7 | 0.41 |
| Regensburg | DEU | 155006 | 79 | 10.8 | 0.43 |
| Reggiodicalabria | ITA | 159399 | 132 | 43.9 | 0.37 |
| Reggionellemilia | ITA | 166323 | 233 | 29.4 | 0.92 |
| Regina | CAN | 222523 | 144 | 16.6 | 0.9 |
| Reims | FRA | 206587 | 79 | 14.9 | 0.54 |
| Remscheid | DEU | 109074 | 73 | 12.9 | 0.79 |
| Rennes | FRA | 401231 | 50 | 21.6 | 0.1 |
| Reus | ESP | 118818 | 52 | 14.6 | 0.55 |
| Reutlingen | DEU | 114364 | 86 | 12.6 | 0.47 |
| Richland | USA | 429557 | 1938 | 52.9 | 2.54 |

Table 1: Cities' metrics: population, area, proximity time, and annual $CO_2$ emissions for transport per capita

| City | Country | Population | A (km$^2$) | $s$ (min) | $C_{pc}$ (t) |
|---|---|---|---|---|---|
| Richmond (greater) | USA | 196463 | 160 | 22.7 | 2.26 |
| Riga | LVA | 639193 | 302 | 20.1 | 0.1 |
| Rimini | ITA | 133671 | 135 | 30.1 | 0.92 |
| Roanoke | USA | 94178 | 110 | 33.3 | 1.76 |
| Rochester (mn) | USA | 154508 | 1666 | 39.2 | 6.1 |
| Rochester (ny) | USA | 705122 | 1715 | 35.8 | 2.37 |
| Rock | USA | 156828 | 1860 | 51.8 | 5.76 |
| Rome | ITA | 2681049 | 1239 | 13.6 | 0.63 |
| Rostock | DEU | 205177 | 170 | 17.9 | 0.51 |
| Rotterdam | NLD | 1287079 | 651 | 28.7 | 0.85 |
| Rouen | FRA | 487666 | 108 | 29.0 | 0.17 |
| Ruhr | DEU | 3636034 | 1887 | 11.6 | 0.84 |
| Rushmoor | GBR | 92557 | 38 | 16.0 | 0.62 |
| Saarbrucken | DEU | 176615 | 169 | 11.6 | 1.26 |
| Sacramento | USA | 2252266 | 7994 | 40.3 | 1.83 |
| Saginaw | USA | 177878 | 2094 | 73.6 | 4.58 |
| Saintbrieuc | FRA | 114881 | 270 | 35.4 | 0.55 |
| Saintetienne | FRA | 371311 | 558 | 26.3 | 0.66 |
| Saintnazaire | FRA | 109789 | 217 | 44.2 | 0.5 |
| Salamanca | ESP | 132764 | 38 | 7.8 | 0.35 |
| Salerno | ITA | 123967 | 59 | 35.6 | 0.34 |
| Salt lake | USA | 1513628 | 2647 | 22.4 | 1.43 |
| San antonio | USA | 2000475 | 3081 | 38.7 | 1.59 |
| San diego | USA | 3227885 | 8484 | 31.1 | 1.66 |
| San francisco (greater) | USA | 5944926 | 7694 | 18.6 | 1.4 |
| San joaquin | USA | 799145 | 3372 | 47.1 | 2.21 |
| Sangamon | USA | 189398 | 2213 | 43.7 | 5.24 |

Table 1: Cities' metrics: population, area, proximity time, and annual $CO_2$ emissions for transport per capita

| City | Country | Population | A (km$^2$) | $s$ (min) | $C_{pc}$ (t) |
|---|---|---|---|---|---|
| Santa barbara | USA | 427005 | 4437 | 39.7 | 1.94 |
| Santa cruz | USA | 240941 | 1130 | 26.7 | 2.24 |
| Santander | ESP | 212609 | 68 | 9.4 | 0.41 |
| Santiago de compostela | ESP | 87853 | 220 | 16.0 | 1.29 |
| Sapporo | JPN | 1909233 | 1184 | 15.0 | 0.38 |
| Saragossa | ESP | 664638 | 898 | 23.0 | 0.75 |
| Sarasota | USA | 799735 | 3198 | 55.5 | 1.84 |
| Saskatoon | CAN | 269606 | 217 | 20.8 | 0.87 |
| Sassari | ITA | 121375 | 545 | 34.0 | 1.0 |
| Schwerin | DEU | 90008 | 124 | 16.6 | 0.67 |
| Scott | USA | 297613 | 2341 | 51.8 | 4.39 |
| Seattle | USA | 3716435 | 13303 | 29.1 | 1.7 |
| Sebastian | USA | 121133 | 1216 | 64.7 | 3.16 |
| Sedgwick | USA | 513406 | 2592 | 56.0 | 2.63 |
| Sendai | JPN | 1186152 | 861 | 15.5 | 0.59 |
| Seoul | KOR | 21596984 | 3302 | 15.0 | 0.48 |
| Seville | ESP | 873772 | 560 | 14.6 | 0.54 |
| Shawnee | USA | 172756 | 1408 | 38.2 | 5.3 |
| Sheffield | GBR | 1103652 | 983 | 16.5 | 0.99 |
| Sherbrooke | CAN | 167388 | 364 | 27.6 | 1.69 |
| Shimonoseki | JPN | 237927 | 706 | 43.6 | 1.28 |
| Shizuoka | JPN | 666357 | 1289 | 26.8 | 0.82 |
| Shunan | JPN | 180195 | 725 | 44.2 | 2.0 |
| Siauliai | LTU | 82544 | 77 | 15.9 | 0.24 |
| Siegen | DEU | 99127 | 114 | 14.0 | 1.02 |
| Siracusa | ITA | 111927 | 205 | 25.7 | 1.27 |
| Slough | GBR | 158823 | 32 | 19.2 | 0.37 |

Table 1: Cities' metrics: population, area, proximity time, and annual $CO_2$ emissions for transport per capita

| City | Country | Population | A (km$^2$) | $s$ (min) | $C_{pc}$ (t) |
|---|---|---|---|---|---|
| Solingen | DEU | 157491 | 89 | 9.6 | 0.69 |
| Sonoma | USA | 475316 | 3742 | 39.2 | 2.63 |
| Southampton | GBR | 388727 | 130 | 11.7 | 0.68 |
| Spokane | USA | 482919 | 4467 | 34.8 | 2.91 |
| St catharines | CAN | 134473 | 98 | 18.5 | 1.37 |
| St johns | CAN | 116833 | 373 | 23.8 | 2.41 |
| Stanislaus | USA | 567139 | 3107 | 51.2 | 1.61 |
| Stark | USA | 343460 | 1494 | 48.7 | 2.95 |
| Stevenage | GBR | 86029 | 25 | 14.5 | 0.29 |
| Stockholm | SWE | 1091602 | 783 | 12.9 | 0.24 |
| Stoke-on-trent | GBR | 380793 | 303 | 19.9 | 0.78 |
| Strasbourg | FRA | 468977 | 134 | 19.1 | 0.36 |
| Stuttgart | DEU | 881210 | 349 | 9.1 | 0.5 |
| Summit | USA | 506281 | 1084 | 38.0 | 2.74 |
| Sumter | USA | 150098 | 1326 | 155.1 | 3.24 |
| Sunderland | GBR | 271195 | 139 | 16.0 | 0.85 |
| Sutter | USA | 108824 | 1424 | 75.0 | 3.83 |
| Swansea | GBR | 253006 | 378 | 22.8 | 1.28 |
| Swindon | GBR | 240864 | 230 | 16.3 | 0.93 |
| Sydney | AU | 4687592 | 3654 | 19.4 | 0.65 |
| Szeged | HUN | 131179 | 282 | 20.7 | 1.09 |
| Szekesfehervar | HUN | 92267 | 169 | 21.6 | 0.71 |
| Takamatsu | JPN | 391252 | 369 | 20.8 | 0.68 |
| Takasaki | JPN | 791107 | 961 | 22.8 | 1.29 |
| Talavera de la reina | ESP | 97838 | 184 | 22.9 | 0.71 |
| Tallahassee | USA | 233118 | 1772 | 33.7 | 3.4 |
| Tallinn | EST | 362097 | 155 | 11.6 | 0.39 |

Table 1: Cities' metrics: population, area, proximity time, and annual $CO_2$ emissions for transport per capita

| City | Country | Population | A (km$^2$) | $s$ (min) | C$_{pc}$ (t) |
|---|---|---|---|---|---|
| Tampa-hernando | USA | 215039 | 1199 | 110.4 | 2.2 |
| Tampa-hillsborough | USA | 1443009 | 2595 | 40.5 | 1.63 |
| Tampa-pinellas | USA | 867981 | 830 | 27.8 | 1.01 |
| Taranto | ITA | 150761 | 238 | 38.5 | 0.52 |
| Tarragona | ESP | 139828 | 55 | 22.1 | 0.4 |
| Tartu | EST | 91330 | 152 | 14.8 | 0.45 |
| Tauranga | NZL | 132798 | 135 | 22.6 | 0.36 |
| Taylor | USA | 124926 | 2197 | 41.8 | 5.5 |
| Telford and wrekin | GBR | 175094 | 289 | 21.2 | 0.93 |
| Terni | ITA | 107980 | 210 | 36.0 | 0.57 |
| Terrebonne | USA | 114395 | 852 | 119.4 | 2.41 |
| Thanet | GBR | 139967 | 103 | 13.6 | 0.38 |
| Thurston | USA | 291264 | 1847 | 53.5 | 2.34 |
| Tokushima | JPN | 286519 | 210 | 24.9 | 0.69 |
| Tokyo | JPN | 34597500 | 7192 | 12.5 | 0.29 |
| Toledo | ESP | 92622 | 218 | 23.9 | 1.38 |
| Tomakomai | JPN | 153132 | 553 | 36.8 | 1.25 |
| Torbay | GBR | 131577 | 63 | 15.2 | 0.48 |
| Torino | ITA | 917345 | 152 | 14.4 | 0.26 |
| Toronto | CAN | 6838851 | 3244 | 14.5 | 0.68 |
| Torrevieja | ESP | 94229 | 66 | 14.4 | 0.49 |
| Toulon | FRA | 407036 | 144 | 28.1 | 0.24 |
| Toulouse | FRA | 718312 | 206 | 20.8 | 0.42 |
| Tours | FRA | 281019 | 131 | 20.7 | 0.58 |
| Toyama | JPN | 399280 | 1046 | 31.0 | 1.62 |
| Toyohashi | JPN | 562101 | 416 | 22.6 | 0.72 |
| Trento | ITA | 117732 | 172 | 26.2 | 0.93 |

Table 1: Cities' metrics: population, area, proximity time, and annual $CO_2$ emissions for transport per capita

| City | Country | Population | A (km$^2$) | $s$ (min) | $C_{pc}$ (t) |
|---|---|---|---|---|---|
| Treviso | ITA | 91042 | 55 | 24.8 | 0.59 |
| Trier | DEU | 120307 | 115 | 13.0 | 0.87 |
| Trieste | ITA | 198858 | 84 | 19.8 | 0.35 |
| Trois rivieres | CAN | 141451 | 283 | 26.6 | 1.45 |
| Troyes | FRA | 126397 | 42 | 18.1 | 0.2 |
| Tubingen | DEU | 88913 | 107 | 10.4 | 0.72 |
| Tulare | USA | 499535 | 9638 | 86.0 | 3.14 |
| Tulsa | USA | 611000 | 1470 | 32.0 | 2.63 |
| Tuscaloosa | USA | 198793 | 3181 | 82.5 | 4.34 |
| Ube | JPN | 153879 | 281 | 38.7 | 1.04 |
| Udine | ITA | 105740 | 56 | 19.7 | 0.6 |
| Ulm | DEU | 182149 | 199 | 12.3 | 0.87 |
| Uppsala | SWE | 226312 | 2226 | 30.9 | 0.72 |
| Usti nad labem | CZE | 90241 | 93 | 24.1 | 0.72 |
| Utah | USA | 693699 | 4623 | 28.3 | 1.51 |
| Utsunomiya | JPN | 495594 | 415 | 20.8 | 0.74 |
| Valence | FRA | 124663 | 62 | 29.3 | 0.32 |
| Valencia | ESP | 962540 | 230 | 12.3 | 0.37 |
| Valenciennes | FRA | 195080 | 58 | 37.1 | 0.31 |
| Valladolid | ESP | 290786 | 195 | 15.6 | 0.48 |
| Vancouver | CAN | 2410850 | 1429 | 12.9 | 0.73 |
| Vanderburgh | USA | 176034 | 608 | 50.8 | 2.08 |
| Vannes | FRA | 124016 | 450 | 37.4 | 0.78 |
| Varese | ITA | 87565 | 53 | 25.1 | 0.33 |
| Vasteras | SWE | 145973 | 1031 | 33.1 | 0.88 |
| Venezia | ITA | 188793 | 122 | 29.2 | 0.82 |
| Ventura | USA | 863234 | 3462 | 34.5 | 1.78 |

Table 1: Cities' metrics: population, area, proximity time, and annual $CO_2$ emissions for transport per capita

| City | Country | Population | A (km$^2$) | $s$ (min) | $C_{pc}$ (t) |
|---|---|---|---|---|---|
| Verona | ITA | 263763 | 198 | 23.6 | 0.78 |
| Verviers (greater city) | BEL | 110311 | 286 | 18.9 | 2.3 |
| Vicenza | ITA | 125387 | 79 | 23.8 | 0.79 |
| Victoria | CAN | 240588 | 143 | 13.9 | 0.81 |
| Vienna | AUT | 1706117 | 411 | 10.0 | 0.4 |
| Vigo | ESP | 263566 | 107 | 15.8 | 0.38 |
| Villingen-schwenningen | DEU | 82683 | 165 | 12.9 | 0.97 |
| Vilnius | LTU | 526328 | 399 | 21.5 | 0.34 |
| Virginia beach | USA | 995459 | 1605 | 38.3 | 1.82 |
| Vitoria | ESP | 238569 | 275 | 14.5 | 0.79 |
| Volusia-daytona beach | USA | 525425 | 2740 | 60.0 | 2.21 |
| Wakayama | JPN | 379593 | 241 | 21.0 | 0.45 |
| Wake | USA | 1180179 | 2205 | 37.0 | 1.7 |
| Warrington | GBR | 213599 | 181 | 14.5 | 1.56 |
| Warsaw | POL | 1697605 | 515 | 13.1 | 0.49 |
| Washington (greater) | USA | 6597578 | 7986 | 26.4 | 1.62 |
| Washoe | USA | 494598 | 8316 | 50.9 | 2.52 |
| Washtenaw | USA | 342858 | 1861 | 38.2 | 3.42 |
| Waveney | GBR | 117630 | 373 | 23.6 | 0.94 |
| Webb | USA | 305935 | 7390 | 63.8 | 2.43 |
| Weber | USA | 258410 | 1271 | 28.9 | 1.36 |
| Weld | USA | 330009 | 8760 | 74.9 | 3.35 |
| Wellington | NZL | 368709 | 298 | 16.5 | 0.39 |
| West midlands urban area | GBR | 2760751 | 912 | 13.8 | 0.64 |
| Whatcom | USA | 219830 | 3443 | 50.0 | 3.48 |
| Wichita | USA | 120824 | 1461 | 66.4 | 4.78 |
| Wiesbaden | DEU | 289855 | 202 | 11.6 | 0.88 |

Table 1: Cities' metrics: population, area, proximity time, and annual $CO_2$ emissions for transport per capita

| City | Country | Population | A (km$^2$) | $s$ (min) | $C_{pc}$ (t) |
|---|---|---|---|---|---|
| Windsor | CAN | 207486 | 145 | 22.0 | 1.13 |
| Winnebago (il) | USA | 296623 | 1329 | 43.7 | 2.13 |
| Winnebago (wi) | USA | 166131 | 1222 | 39.2 | 3.91 |
| Winnipeg | CAN | 725739 | 472 | 18.6 | 0.79 |
| Winterthur | CHE | 112507 | 67 | 8.2 | 0.52 |
| Woodbury | USA | 110871 | 2903 | 52.3 | 7.21 |
| Worcester | GBR | 103325 | 33 | 13.2 | 0.23 |
| Worthing | GBR | 110709 | 32 | 10.3 | 0.25 |
| Wrexham | GBR | 141417 | 499 | 31.1 | 1.7 |
| Wuppertal | DEU | 344564 | 169 | 11.1 | 0.71 |
| Wurzburg | DEU | 125637 | 86 | 11.6 | 0.64 |
| Wycombe | GBR | 181258 | 324 | 19.1 | 1.37 |
| Yakima | USA | 237812 | 8809 | 77.1 | 5.2 |
| Yamagata | JPN | 234847 | 388 | 25.2 | 1.08 |
| Yellowstone | USA | 157740 | 4794 | 59.7 | 5.45 |
| Yokkaichi | JPN | 779940 | 1114 | 32.5 | 1.04 |
| Yonago | JPN | 138428 | 135 | 19.0 | 0.73 |
| York | GBR | 215660 | 270 | 14.4 | 0.73 |
| Yuma | USA | 222411 | 6282 | 89.8 | 3.08 |
| Zurich | CHE | 364308 | 85 | 9.4 | 0.44 |
| Zwickau | DEU | 89377 | 103 | 15.0 | 0.65 |



\end{document}